\documentclass[12pt,a4paper]{article}
\usepackage{a4wide}
\usepackage{epsfig}
\usepackage{amsmath}
\usepackage{amssymb}
\usepackage{cite}

\usepackage{multirow} 
\usepackage{wasysym}  

\def\baselinestretch{1.2}
\def    \beq           	{\begin{equation}}
\def    \eeq           	{\end{equation}}
\def    \bea           	{\begin{eqnarray}}
\def    \eea           	{\end{eqnarray}}
\def    \nn            	{\nonumber}  
\def    \raw           	{\rightarrow}

\def    \lraw           {\leftrightarrow}

\def	\tr		{{\rm tr}}
\def	\Tr		{{\rm Tr}}

\def	\cN		{{\mathcal{N}}}
\def	\cL		{{\mathcal{L}}}
\def	\cW		{{\mathcal{W}}}
\def	\cO		{{\mathcal{O}}}
\def 	\bZ		{{\mathbb Z}}

\def	\bR		{{\mathbb R}}

\newcommand{\drawsquare}[2]{\hbox{%
\rule{#2pt}{#1pt}\hskip-#2pt
\rule{#1pt}{#2pt}\hskip-#1pt
\rule[#1pt]{#1pt}{#2pt}}\rule[#1pt]{#2pt}{#2pt}\hskip-#2pt
\rule{#2pt}{#1pt}}
\newcommand{\fund}{\raisebox{-.5pt}{\drawsquare{6.5}{0.4}}}

\newcommand{\Yasym}{\raisebox{-3.5pt}{\drawsquare{6.5}{0.4}}\hskip-6.9pt%
        \raisebox{3pt}{\drawsquare{6.5}{0.4}}}

\newcommand{\id}{1\!\!1}

\def	\cK		{{\cal{K}}}
\def	\cA		{{\cal{A}}}
\def	\cM		{{\cal{M}}}
\def	\cR		{{\cal{R}}}
\def	\ctK		{\tilde{\cal{K}}}
\def	\ctA		{\tilde{\cal{A}}}
\def	\ctM		{\tilde{\cal{M}}}
\def	\ctL		{\tilde{\cal{L}}}

\def	\OR		{{\Omega \cal{R}}}
\def 	\ba		{{\mathbf{a}}}
\def 	\bb		{{\mathbf{b}}}
\def 	\b1		{{\mathbf{1}}}

\newcommand{\ket}[1]{| \, {#1} \, \rangle}

\newcommand{\half}{\frac{1}{2}}

\newcommand{\LS}{\ \ \ \ \ \ \ \ \ \ }
\newcommand{\ls}{\ \ \ \ \ }
\newcommand{\bsubeq}{\begin{subequations}}
\newcommand{\esubeq}{\end{subequations}}
\newcommand{\noi}{\noindent}
\newcommand{\slb}{\scalebox}

\newcommand{\Z}{\mathbb Z}

\usepackage{color}

\newcommand{\vol}{{\rm Vol}}
\newcommand{\inv}{{\rm inv}}

\makeatletter

 \@addtoreset{equation}{section}
\makeatother

\renewcommand{\thefootnote}{\fnsymbol{footnote}}

\begin{document}

\allowdisplaybreaks{
\thispagestyle{empty}

\begin{titlepage}

\begin{flushright}
KUNS-2100 \\
YITP-07-82 \\
December 2007 
\end{flushright}

\vspace{10mm}

\begin{center}
\slb{1.8}{Type IIA orientifolds and orbifolds}

\slb{1.8}{on non-factorizable tori}

\vspace{15mm}

\slb{1.1}{Tetsuji Kimura$^{1*}$, \ 
Mitsuhisa Ohta$^{1\dagger}$ 
\ and \  
Kei-Jiro Takahashi$^{2\ddagger}$ 
}

\vspace{5mm}

$^{1}${\sl \normalsize
Yukawa Institute for Theoretical Physics, Kyoto University,
Kyoto 606-8502, Japan} 

$^2${\sl \normalsize
Department of Physics, Kyoto University,
Kyoto 606-8502, Japan} 

\vspace{7mm}

{
\begin{tabular}{c}
$^*${\tt tetsuji@yukawa.kyoto-u.ac.jp} \\
$^{\dagger}${\tt mituhisa@yukawa.kyoto-u.ac.jp} \\
$^{\ddagger}${\tt keijiro@gauge.scphys.kyoto-u.ac.jp}
\end{tabular}
}

\vspace{15mm}


\begin{abstract}
We investigate
Type II orientifolds on non-factorizable torus with and without
its oribifolding. 
We explicitly calculate the Ramond-Ramond tadpole from string
one-loop amplitudes, and confirm that the consistent number of
orientifold planes is directly derived from the Lefschetz fixed point theorem. 
We furthermore classify 
orientifolds on non-factorizable ${\mathbb Z}_N \times {\mathbb Z}_M$ orbifolds, 
and construct new supersymmetric Type IIA orientifold models on them.
\end{abstract}

\end{center}

\end{titlepage}

\renewcommand{\thefootnote}{\arabic{footnote}}
\setcounter{footnote}{0}

%

\newpage

\section{Introduction}

Many  attempts have been made for constructing string vacua 
using D-branes in order to realize the Standard Model. 
In Type IIA orientifolds, 
intersecting D-brane models provide chiral spectra
\cite{Aldazabal:2000dg, Ibanez:2001nd, Blumenhagen:2001te,
  Bailin:2001ie,Kokorelis:2002zz, Higaki:2005ie, Blumenhagen:2002gw,
  Blumenhagen:2002wn, Cvetic:2001tj, Cvetic:2001nr, Kokorelis:2004gb,Honecker:2003vq,
  Cvetic:2006by},
which feature some of the 
properties of the supersymmetric or non-supersymmetric Standard Model 
(see
\cite{Blumenhagen:2005mu, MarchesanoBuznego:2003hp, Blumenhagen:2006ci}
and references therein).
We know now vast number of perturbative vacua in the landscape 
of string theory, 
and it is of great importance to further investigate 
possible vacua in string theory construction
\cite{Susskind:2003kw, Dijkstra:2004cc, Lebedev:2006kn, Forste:2007zb}.

Most of Type IIA models compactified on six-dimensional spaces 
have been constructed by orbifold tori given
by ${\mathbb Z}_N$ 
\cite{Blumenhagen:2002gw,Blumenhagen:2002wn}, 
${\mathbb Z}_2 \times {\mathbb Z}_2$ 
\cite{Berkooz:1996dw,Cvetic:2001tj, Cvetic:2001nr} 
and ${\mathbb Z}_4 \times {\mathbb Z}_2$
\cite{Honecker:2003vq,Cvetic:2006by}, 
whose point group is defined by the Coxeter elements.
In the case of $\Z_N$ orbifold, some models compactified 
on non-factorizable tori 
are constructed by Coxeter elements \cite{Blumenhagen:2004di}, 
while, in the case of ${\mathbb Z}_N \times {\mathbb Z}_M$
Coxeter orbifolds \cite{Forste:2000hx}, 
the compact spaces are factorized to 
$T^2 \times T^2 \times T^2$.
Recently, however, 
non-factorizable ${\mathbb Z}_2 \times {\mathbb Z}_2$ orbifolds 
were constructed 
in heterotic string \cite{Donagi:2004ht, Faraggi:2006bs, Forste:2006wq}, 
and in Type IIA string \cite{Forste:2007zb}. 
One of the authors in this paper
recently classified ${\mathbb Z}_N \times {\mathbb Z}_M$ orbifold models 
on non-factorizable tori \cite{Takahashi:2007qc,Takahashi:2007vs}. 
Non-factorizable orbifolds possess different geometries from
factorizable ones 
because the number of fixed tori, and the Euler numbers,
in six-dimensional
spaces can be less than those of the factorizable ones.
Such non-factorizable orbifolds can be applied to Type IIA string models, 
which give rise to rather richer structure by inclusion of D-branes. 

For the consistency of theory the tadpole cancellation is required 
(see \cite{Gimon:1996rq,Aldazabal:1998mr,Blumenhagen:1999md,Blumenhagen:2000vk,
  Blumenhagen:2000ea, Blumenhagen:2001te},  
and for review, \cite{Dabholkar:1997zd,Angelantonj:2002ct,Ott:2003yv}).
We explicitly calculate string one-loop amplitudes on the 
Klein bottle, the annulus and the M\"obius strip 
on non-factorizable tori and orbifolds,
and confirm that the consistent number of
orientifold planes (O-planes) 
is directly derived from the Lefschetz fixed point theorem 
via the cancellations of Ramond-Ramond (RR) tadpole. 
We give a systematic way to construct various models on non-factorizable orbifolds.
Interestingly, we further find 
new feature of non-factorizable $\Z_2 \times \Z_2$ orbifolds,
in which the numbers of O-planes depend on three-cycles. 

This paper is organized as follows:
In section \ref{ONFT} we describe the tadpole cancellation condition 
on generic non-factorizable tori. 
In this analysis
the Lefschetz fixed point theorem makes
the cancellation condition simplified and provides an intuitive picture.
We apply this formula to orientifold models which have been already
well-investigated. 
In section \ref{ZZorient}
we explicitly construct Type IIA orientifolds on 
${\mathbb Z}_4 \times {\mathbb Z}_2$ and ${\mathbb Z}_2 \times {\mathbb Z}_2$ 
orbifolds on the $D_6$ Lie root lattice.
Because the contributions of untwisted sector in orbifolds 
are given by the same forms of those in tori, 
the formula, which is derived from the Lefschetz fixed point theorem, 
provides a necessary condition on non-factorizable orbifolds. 
We describe general features of orientifold constructions 
on non-factorizable orbifolds.  
Section \ref{Conclusion} is devoted to the conclusion.
In appendix \ref{App-LRL} we explain details of 
the classification of orientifolds and orbifolds on the Lie root lattices. 
In appendix \ref{App-lattice} we summarize a set of useful conventions
to describe non-factorizable tori in terms of the lattice space and
its dual. 
In appendix \ref{App-1loopPF} we briefly review the string
one-loop amplitudes which are given by the Klein bottle, the annulus
and the M\"{o}bius strip as the worldsheet topologies.


\section{Orientifold on non-factorizable torus}
\label{ONFT}

In this section we will evaluate RR-tadpole cancellation conditions of torus
compactification in Type IIA string theory in the presence of
D6-branes and orientifold planes (O6-planes).
We will show a method to analyze the orientifold models on
non-factorizable tori, which can be applied to any kind of
torus compactifications.
We introduce a set of general formula 
for the tadpole amplitudes in RR-sector
on non-factorizable tori, 
which are defined by the Lie root lattices. 
Utilizing the Lefschetz fixed point theorem,
we can check the tadpole cancellation condition not only on the
usual factorizable tori but also on the non-factorizable ones in a
quite simple way.
We will further apply this method to orbifold models in section \ref{ZZorient}.

\subsection{RR-tadpole and the Lefschetz fixed point theorem}

We consider the Type IIA models compactified on a six-torus
$T^6$. 
A six-torus could be regarded as a six-dimensional Euclidean space
$\bR^6$  divided by a lattice $\Lambda$, i.e., $T^6 = \bR^6/\Lambda$.
As we will see, the structure of the lattice $\Lambda$ plays 
a central role in the analysis of this paper. 
Here let us consider orientifolds in Type IIA given in the following way:
\beq
\frac{\text{Type IIA on $T^6$}}{\Omega \cR},
\label{orientifold}
\eeq
where $\Omega$ is the worldsheet parity operator,
and $\cR$ is the orientifold involution which 
indicates the reflection of three directions in $T^6$. 
Usually the action $\cR$ can be given as
\beq
\cR:~ z_i \raw \bar{z}_i.
\label{C-type}
\eeq
In order to construct consistent effective theories in
four-dimensional spacetime, 
we study the tadpole cancellation condition in the presence of orientifolds.
The tadpole amplitude is derived from 
the string one-loop graphs whose topologies are the
Klein bottle, the annulus, and the M\"{o}bius strip. 
These amplitudes are represented as ${\cal K}$, 
${\cal A}$ and ${\cal M}$, respectively.
Here let us explicitly describe their amplitudes in terms of a
modulus $t$ in the loop channel as follows:
\bsubeq
\begin{align}
{\cal K} &= 4c\int_0^\infty \frac{dt}{t^3} 
\Tr_{\text{closed}}\left(
  \frac{\Omega {\cal R}}{2}\left(\frac{1+\left(-1\right)^{F}}{2}\right) 
(-1)^{\mathbf s}
e^{-2\pi t\left( L_0+\bar{L}_0\right)}\right),
   \label{kb1} \\
{\cal A} &= c \int_0^\infty
  \frac{dt}{t^3} \Tr_{\text{open}}\left( \frac{1}{2}
  \left(\frac{1 +\left(-1\right)^{F}}{2}\right)
(-1)^{\mathbf s}
e^{-2\pi t L_0}\right) 
, \label{an1} \\
{\cal M} &= c \int_0^\infty
  \frac{dt}{t^3} \Tr_{\text{open}}\left(\frac{\Omega {\cal
  R}}{2}
  \left(\frac{1 
  +\left(-1\right)^{F}}{2}\right)
(-1)^{\mathbf s}
e^{-2\pi t L_0}\right) 
, \label{ms1}
\end{align}
\esubeq
where $F$ and ${\mathbf s}$ denote the fermion numbers in the worldsheet
and in the spacetime, respectively;
the overall coefficient $c$ is given by
$c \equiv V_4 / (8\pi^2 {\alpha}^\prime)^2$, 
where $V_4$ is from the integration over momenta in non-compact
directions. 
Since the divergence from the RR-tadpole should
be evaluated in the tree channel, 
which is described by the $l$-modulus,
we should rewrite them via the modular transformation,
even though 
the computations of the amplitudes are easier in the loop
channel given by $t$-modulus. 
The RR-sectors in the tree channel 
which we should evaluate in order to see the tadpole
cancellation in the presence of orientifold planes and D-branes,
correspond to the states with the following insertions 
in the loop channel \cite{Polchinski:1987tu}:
\begin{align}
\text{Klein bottle} &: \text{closed string, NS-NS sector, $(-1)^F$} \nn \\
\text{annulus} &: \text{open string, R sector}\\
\text{M\"obius strip} &: \text{open string, NS sector, $(-1)^F$} \nn 
\end{align}
In this paper we calculate these amplitudes 
for the case cases that D-branes are parallel on O-planes. 
Then the amplitudes can be written in the form  
as follows:
\bsubeq \label{eta-L-KAMamp}
\begin{align}
\cK &= 
c (1_{\text{RR}}-1_{\text{NSNS}}) \int_0^\infty \frac{dt}{t^3} 
\frac{\vartheta \left[ 0 \atop 1/2 \right]^4 
}{\eta^{12}} \cL_{\cK},\label{K-torus} \\
\cA &= 
\frac{c}{4} (1_{\text{RR}}-1_{\text{NSNS}}) 
\{ (\tr(\gamma_1))^2 \} \int_0^\infty \frac{dt}{t^3} 
\frac{\vartheta \left[ 0 \atop 1/2 \right]^4
}{\eta^{12}} \cL_{\cA},\label{A-torus} \\
\cM &= 
- \frac{c}{4} (1_{\text{RR}}-1_{\text{NSNS}}) 
\{ \tr(\gamma_\OR^{-1}\gamma_\OR^T) \}
\int_0^\infty \frac{dt}{t^3} 
\frac{\vartheta \left[ 1/2 \atop 0 \right]^4
}{\eta^{12}} \cL_{\cM}, \label{M-torus} 
\end{align}
\esubeq
where the string oscillation modes are represented with respect to 
the $\vartheta$-function
and the Dedekind $\eta$-function, while the zero modes are
given by $\cL_{\cK}$, $\cL_{\cA}$ and $\cL_{\cM}$. 
The $\gamma$ matrices are orientifold actions on the Chan-Paton factors
in the notation of \cite{Gimon:1996rq}.
Due to the spacetime supersymmetry, 
the total amplitudes from RR- and NSNS-sectors should be cancelled to
each other, 
as seen the factor $(1_{\text{RR}}-1_{\text{NSNS}})$ on each amplitude
in (\ref{eta-L-KAMamp}). 
The mapping between the two different moduli $t$ and $l$ in these
channels is also given as
\beq
\text{Klein bottle}:~ t = \frac{l}{4}
, \ls 
\text{annulus}:~ t = \frac{l}{2} 
, \ls 
\text{M\"{o}bius strip}:~ t = \frac{l}{8} 
. \label{tl-KAM}
\eeq
To evaluate the RR-tadpole generated by the orientifold, 
we extract only the contributions 
from RR-sector in the tree channel. 
In the IR limit $l \raw \infty$ 
the divergence from the RR-tadpole should be cancelled, 
\beq
{\ctK}_{\text{RR}}+{\ctA}_{\text{RR}}+{\ctM}_{\text{RR}} \raw 0,
\label{RR-cancel}
\eeq
where $\ctK_{\text{RR}}$, $\ctA_{\text{RR}}$ and $\ctM_{\text{RR}}$
are RR-tadpole contributions in the tree channel mapped from $\cK$, $\cA$ and
$\cM$ in the loop channel under the modular transformation, respectively.

Now let us evaluate the zero mode contributions 
$\cL_{\cK, \cA, \cM}$ in (\ref{eta-L-KAMamp}) given by the momentum modes and
the winding modes. 
${\bf p}$ and winding modes ${\bf w}$
can be written in terms of a set of certain basis vectors 
$\{ {\bf p}_i \}$ and $\{ {\bf w}_i \}$, respectively:
\begin{align}
{\bf p} &= \sum_i n_i {\bf p}_i  
, \ls
{\bf w} = \sum_i m_i {\bf w}_i
, \ls  m_i, n_i \in \bZ
.
\end{align}
The zero mode contribution to the loop channel amplitudes is
\beq
\cL \equiv \sum_{n_i} \exp \Big(-\delta \pi t n_i M_{ij} n_j \Big) 
\cdot \sum_{m_i} \exp \Big(-\delta \pi t m_i W_{ij} m_j \Big) ,
\label{zero-part}
\eeq
where $n_i$, $m_i \in \bZ$, 
$M_{ij} = {\bf p}_i \cdot {\bf p}_j$, 
$W_{ij} = {\bf w}_i \cdot {\bf w}_j$ and 
$\delta = 1$ for Klein bottle, 
$\delta = 2$ for annulus and M\"obius strip. 
Using the generalized Poisson resummation formula, 
we can rewrite
\beq
\sum_{n_i} \exp \Big(-\pi t n_i A_{ij} n_j \Big)
=
\frac{1}{t^{\frac{{\rm dim}(A)}{2}} (\det A)^\frac12} 
\sum_{n_i} \exp \Big( - \frac{\pi}{t} n_i A^{-1}_{ij} n_j \Big)
.
\eeq
When we move to the tree channel by using (\ref{tl-KAM}),
the zero mode contribution $\cL$ is 
\begin{align}
\cL
&=
\sum_{n_i} \frac{\left(\frac{{\alpha} l}{\delta} \right)^3}{\sqrt{\det M \det W}}
\exp \Big(-\pi \frac{{\alpha} l}{\delta} t n_i M^{-1}_{ij} n_j \Big)
\cdot \sum_{m_i} 
\exp \Big( -\pi \frac{{\alpha} l}{\delta} m_i W^{-1}_{ij} m_j \Big) 
, 
\end{align}
which goes to
$\frac{(\frac{{\alpha} l}{\delta})^3}{\sqrt{\det M \det W}}$
in the IR limit $l \raw \infty$.

We consider a six-torus $T^6$ on a lattice $\Lambda$. 
Then different two points in $T^6$ are identified in terms of the
lattice shift vector $r {\bf \alpha}_i \in \Lambda$ as 
\begin{align}
T_{{\bf \alpha}_i} : \ {\bf x} \to {\bf x} + r{\bf \alpha}_i ,
\label{def-torus}
\end{align}
where $r$ is a radius of $T^6$. 
For simplicity we set $r=1$ in the following in this paper . 
Translation operator acting on the momentum states $\ket{{\bf p}}$ is given 
by 
\beq
T_{{\bf \alpha}_i} \ket{{\bf p}} = \exp(2\pi i {\bf p} \cdot {\bf \alpha}_i)
\ket{{\bf p}}
.
\eeq
Then the momentum modes are expressed by dual vector 
${\bf \alpha}^*_i \in \Lambda^*$, 
\beq
{\bf \alpha}_i \cdot {\bf \alpha}^*_j= \delta_{ij}.
\eeq
In the Klein bottle amplitude, 
the momentum modes should be invariant under the action of $\Omega\cR$. 
Thus the vector ${\bf \alpha}_i^*$ consists of 
the $\cR$ invariant sublattice in the dual lattice $\Lambda^*$,
and we have \cite{Forste:2007zb}\footnote{See appendix \ref{App-lattice} 
for the definition of $\Lambda_{\cR,\inv}$ and $\Lambda_{\cR,\perp}$.} 
\beq
\sqrt{\det M^{\cK}} = \vol(\Lambda^*_{\cR,\inv}). 
\eeq
In the same way,
the winding modes ${\bf w}_i$ are given by the lattice vector ${\bf \alpha}_i$ 
invariant under the action $-\cR$ on the lattice space $\Lambda$ 
(with the constant $\alpha^\prime =1$). 
Then  we obtain
\beq
\sqrt{\det W^{\cK}} = \vol(\Lambda_{-\cR,\inv}).
\eeq
One of the simplest way to cancel the RR-tadpole of the O6-plane is 
to add D6-branes parallel to the O6-planes. 
Since the O6-planes lie on the $\cR$ fixed locus,  
the basis vectors which describe three-cycles of the O6-plane are generated from 
$\cR$-invariant sublattice $\Lambda_{\cR,\inv}$. 
Then, in the case of the annulus amplitude, 
the momentum modes are described by the vector in the dual
lattice $(\Lambda_{\cR,\inv})^*$. 
The winding modes are related to the distances between these D6-branes, 
and they are the sublattice projected by $-\cR$, 
i.e., $\Lambda_{-\cR,\perp} \equiv \frac{1-\cR}{2}\Lambda$. 
In the M\"obius strip amplitude the momentum modes are same as 
the ones of the annulus amplitude. 
On the other hand, the winding modes should be 
in the invariant sublattice under $-\Omega\cR$, and it is given by $\Lambda_{-\cR,\inv}$.
Summarizing the above, we obtain the following descriptions:
\bsubeq
\begin{align}
\sqrt{\det M^{\cK}} &= \vol(\Lambda^*_{\cR,\inv}) 
, \label{MK} \\
\sqrt{\det M^{\cA}} =\sqrt{\det M^{\cM}} &=
\vol(\Lambda^*_{\cR,\perp}) 
, \label{MA} \\
\sqrt{\det W^{\cK}} =\sqrt{\det W^{\cM}} &= \vol(\Lambda_{-\cR,\inv})
, \label{WK} \\
\sqrt{\det W^{\cA}} &= \vol(\Lambda_{-\cR,\perp})
, \label{WA} 
\end{align}
\esubeq
where we used the following relations:
\begin{align}
\Lambda^*_{\cR,\perp} &= (\Lambda_{\cR,\inv})^*
, \ls
\vol(\Lambda) 
= \vol(\Lambda_{\cR,\inv}) \cdot \vol(\Lambda_{-\cR,\perp})
.
\end{align}

For the contributions to Chan-Paton factors, 
we have $\gamma_1=\id$ 
so that $\tr(\gamma_1)=N$ is the number of D6-branes. 
Furthermore we require $\gamma_\OR^{-1}\gamma_\OR^T=\id$ 
in order to cancel the RR-tadpole. 

Now we are ready to obtain the RR-tadpole cancellation condition.
The sum of RR-tadpole contributions for large $l$ is asymptotically 
\begin{align}
&\ctK_{\text{RR}} + \ctA_{\text{RR}} + \ctM_{\text{RR}}  \\
&\ls= 
c \int^{\infty} d l \, \left(
\frac{64}{\sqrt{\det M^{\cK} \det W^{\cK}}} 
+ \frac{N^2}{16\sqrt{\det M^{\cA} \det W^{\cA}}}  
- \frac{4N}{\sqrt{\det M^{\cM} \det W^{\cM}}} 
\right) 
\nn \\
&\ls=  
c \int^{\infty} d l \, 
 \frac{1}{16 \vol(\Lambda^*_{\cR,\perp}) \vol(\Lambda_{-\cR,\perp})}
\left( N -4N_{\text{O6}} \right)^2
, 
\label{o6-condition}
\end{align}
where $N_{\text{O6}}$ is the number of the O6-planes 
according to  the Lefschetz fixed point theorem:
\bea
N_{\text{O6}} \equiv \frac{\vol((1-\cR)\Lambda)}{\vol(\Lambda_{-\cR,\inv})}
=2^3 \cdot \frac{\vol(\Lambda_{-\cR,\perp})}{\vol(\Lambda_{-\cR,\inv})}. 
\label{o6-number}
\eea
The equation (\ref{o6-condition}) indicates that 
the RR-tadpole is cancelled 
by D6-branes whose number is four times as many as that of O6-planes. 
Therefore we find that 
it is enough to count the number of O6-planes in (\ref{o6-number})
instead of calculating individual amplitudes. 
For factorizable models, 
we have $\vol(\Lambda_{-\cR,\perp})/\vol(\Lambda_{-\cR,\inv})=1$. 
The condition (\ref{o6-condition}) is also expressed as 
\beq
N \Pi  -4 \Pi_{\text{O6}} =0,
\eeq
where $\Pi$ and $\Pi_{\text{O6}}$ denote three-cycles 
in D6-branes and O6-planes, respectively.

This is the case for O6-planes in Type IIA theory.
We can generalize this tadpole cancellation condition to an
O$q$-plane in type IIA/IIB theory in such a way as
\beq
\left( N -2^{q-4}N_{\text{O$q$}} \right)^2=0,
\eeq
where the number of O$q$-planes is given by
\beq
N_{\text{O$q$}} \equiv \frac{\vol((1-\cR)\Lambda)}{\vol(\Lambda_{-\cR,\inv})}
=2^{9-q} \cdot \frac{\vol(\Lambda_{-\cR,\perp})}{\vol(\Lambda_{-\cR,\inv})}.
\eeq
In the case of an O9-plane,the orientifold action is 
given by $\Omega$, i.e., $\cR= \id$, 
and the above equation is ill-defined, 
however we can calculate it in a same way. 
Then it is appropriate to 
set $\vol(\Lambda_{-\cR,\perp})/\vol(\Lambda_{-\cR,\inv})=1$ for O9-plane.

\subsection{Orientifold models on the Lie root lattices}

Here let us first review the Type IIA orientifold 
on a factorizable torus $T^2 \times T^2 \times T^2$
to fix our notation. 
There are two ways to implement $\OR$ of (\ref{C-type}) 
in each $T^2$.
The lattice $\Lambda_i$ which defines 
the boundary condition of $i$-th $T^2$ is given by 
\beq
\Lambda_i 
= \Big\{ n_{2i-1} {\bf \alpha}_{2i-1} + n_{2i} {\bf \alpha}_{2i} 
\Big| \, n_{2i-1},n_{2i} \in \bZ \Big\} , \ \ \ i=1,2,3 
, 
\eeq
where, for simplicity, we set $r = 1$ in (\ref{def-torus});
${\bf \alpha}_j$ is a simple root of the lattice.
Without loss of generality we can define ${\bf \alpha}_{2i}$ along the 
$x^{2i}$-direction for the orientifold action $\OR$ in (\ref{C-type}), 
which acts crystallographically on the lattice $\Lambda_i$.
Therefore the complex structure $U_i$ on the $i$-th torus $T^2$
should satisfy $\cR U_i = U_i$ modulo the shift given by $\Lambda_i$. 
Then there are only two solutions 
\beq
U_i = ia \ \ \text{or} \ \ \half + ia \; , \ls a\in \bR ,
\eeq
which indicates that  
there are two distinct lattices 
for the $\cR$ action\footnote{By T-dualizing this torus 
this corresponds to $B$-field which is frozen NS-NS closed moduli 
\cite{Blumenhagen:2000ea,MarchesanoBuznego:2003hp}.}. 
The one is called {\bf A}-type lattice \cite{Blumenhagen:1999ev}, 
whose lattice vector is given by
\beq
{\bf \alpha}_1^{\text{\bf A}} = \sqrt{2}{\bf e}_1 , \ \ \
{\bf \alpha}_2^{\text{\bf A}} = \sqrt{2}{\bf e}_2. 
\label{A-lattice}
\eeq
Notice that in this case the complex structure of the torus is given by
$U = ia$.
The other is called {\bf B}-type lattice, which is given by
\beq
{\bf \alpha}_1^{\text{\bf B}} = {\bf e}_1 -{\bf e}_2, \ \ \
{\bf \alpha}_2^{\text{\bf B}} = {\bf e}_1 +{\bf e}_2. 
\label{B-lattice}
\eeq
This corresponds to the case $U = \frac12 + ia$.
We can see it 
by the re-definition of the vector
${\bf \alpha}_2^{\text{\bf B}} \raw 
- {\bf \alpha}_1^{\text{\bf B}}+{\bf \alpha}_2^{\text{\bf B}}$. 
Then we have two distinct theories which depend on the choice of {\bf A}-type or
{\bf B}-type lattices in Figure \ref{ab-torus}. 
For example, the number of fixed loci given by the action of $\cR$ is
two (for the {\bf A}-type) and one (for the {\bf B}-type), 
which associate the total O6-plane charges. 
\begin{figure}[htb]
\begin{center}
\includegraphics{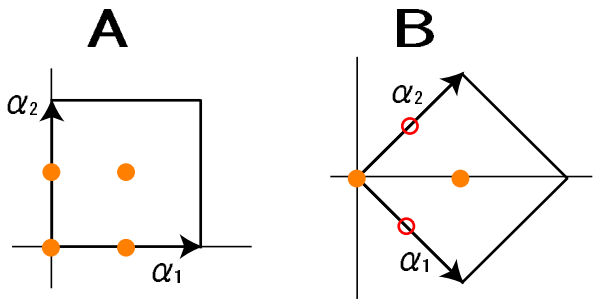}
\caption{\it {\bf A}-type lattice and {\bf B}-type lattice
  in a factorizable torus.}
\label{ab-torus}
\end{center}
\end{figure}
Instead of using the {\bf B}-type lattice, 
we define an equivalent orientifold 
by an alternative definition for $\cR$ on the lattice (\ref{A-lattice}),
\beq
\cR:~ z_j \raw i \bar{z}_j. 
\label{D-type}
\eeq 
In order to distinguish the actions on non-factorizable tori 
from the ones on factorizable torus,
let us attach a label to the action (\ref{D-type}) as {\bf D}, 
and to the one (\ref{C-type}) in the previous subsection 
as {\bf C} \cite{Forste:2007zb}. 
For example we call the models by following $\cR$ action {\bf CCD} model,  
\beq
\cR:~ z_1 \raw \bar{z}_1,~~z_2 \raw \bar{z}_2,~~z_3 \raw i \bar{z}_3.
\eeq
In appendix \ref{App-LRL}, 
we can see that these actions provide convenient tools for 
the classifications of orientifold orbifolds on the Lie root lattices.

First
let us consider the RR-tadpole cancellation conditions in the factorizable models. 
Instead of the direct calculations of 
the zero mode contribution on each $T^2$ and 
of the oscillator modes in the Klein bottle, the annulus and the
M\"{o}bius strip amplitudes, 
it is enough to count the number of O6-planes from (\ref{o6-condition}):
The numbers of O6-planes are 
$N_{\text{O6}} = 8$ (for {\bf AAA}), $4$ (for {\bf AAB}), $2$ (for {\bf ABB}) and 
$1$ (for {\bf BBB}). 
The types of the actions in the $T^2 \times T^2 \times T^2$ are
illustrated in Figure \ref{AAA-AAB}.
\begin{figure}[htb]
\begin{center}
\includegraphics{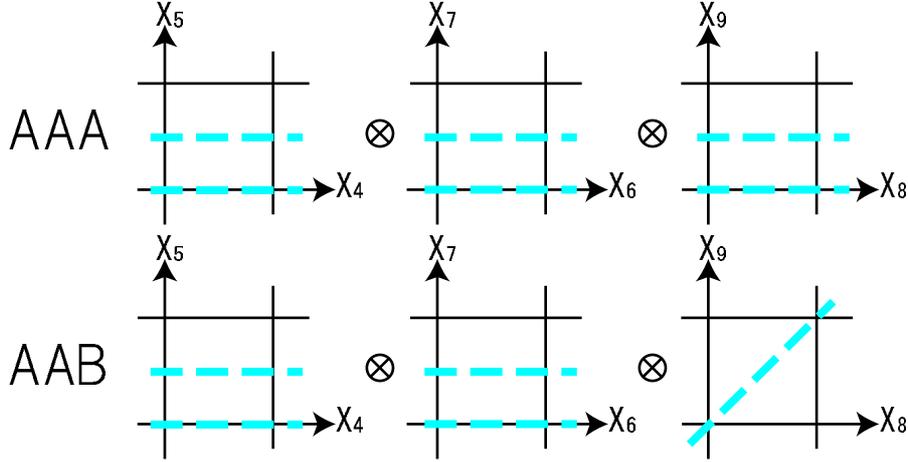}
\caption{\it {\bf AAA} and {\bf AAB} models on a factorizable torus. 
The orientifold planes lie on the dashed blue lines.
In this Figure we used a label {\bf B} 
as the {\bf D}-action on the {\bf A}-lattice.}
\label{AAA-AAB}
\end{center}
\end{figure}
Here we obtain the RR-tadpole cancellation conditions\footnote{Because 
these are the models on factorizable tori,  
and the {\bf C}- and {\bf D}-actions lead to the {\bf A}- and 
{\bf B}-models, respectively.} 
\begin{align}
\begin{array}{l@{\;: \ \ }r}
\text{\bf AAA}& (N-32)^2=0 , \\
\text{\bf AAB}& (N-16)^2=0 ,\\
\text{\bf ABB}& (N-8)^2=0 , \\
\text{\bf BBB}& (N-4)^2=0 .
\end{array}  
\end{align}
These are trivial results which have already been known. 
We emphasize that for the classification of orientifold models 
on non-factorizable tori and orbifolds  
it is convenient to fix the lattices and 
distinguish the models with respect to the definitions of $\cR$.

Next we analyze some typical models
on a non-factorizable\footnote{
In this work a compactified space which cannot be represented as the
direct products of two-torus $T^2$ is called non-factorizable.
For example, six-tori on $D_6$, $A_3 \times A_3$ and $A_3 \times A_2
\times A_1$, 
while six-tori on $A_2 \times A_2 \times A_2$, $A_2 \times D_2 \times
(A_1)^2$ and $(A_1)^6$ are factorizable.} 
tori $T^6$, which 
cannot be expressed as the direct product $T^2 \times T^2 \times T^2$. 
As an example we consider an orientifold model on a non-factorizable torus given
by the Lie root lattice $D_6$. 
In this model the lattice $D_6$ can be given by the simple roots
\begin{align}
{\bf \alpha}_i &= {\bf e}_{i}-{\bf e}_{i+1} 
, \ \ \ 
{\bf \alpha}_6 = {\bf e}_{5} +{\bf e}_{6} , \ \ \ 
 i=1, \dots, 5,
\label{d6-lattice}
\end{align}
where ${\bf e}_i$'s are basis of Cartesian coordinates whose normalization
is given as ${\bf e}_i \cdot {\bf e}_j = \delta_{ij}$.
The orientifold action $\cR$ of the {\bf CCC}-model is 
\beq
\cR: {\bf e}_{2i-1} \raw {\bf e}_{2i-1},~~ 
{\bf e}_{2i} \raw -{\bf e}_{2i},~~i=1,2,3.
\eeq
The number of O6-planes is obtained by means of (\ref{o6-number}). 
In order to evaluate the Lefschetz fixed point theorem, 
we should fix the sublattice spaces $\Lambda_{-\cR,\perp}$
and $\Lambda_{-\cR,\inv}$.
$\Lambda_{-\cR,\perp}$ is a lattice space projected out by $-\cR$, 
and given by
\beq
\Lambda_{-\cR, \perp} =
\Big\{
\sum_{i=1}^3 n_{\perp,i} {\bf \alpha}_{\perp,i} \,\Big|\,n_{\perp,i} \in \Z \Big\},
\eeq
whose basis vectors are given by 
\begin{align}
{\bf \alpha}_{\perp,1} = {\bf e}_{2} , \ \ \ 
{\bf \alpha}_{\perp,2} = {\bf e}_{4} , \ \ \ 
{\bf \alpha}_{\perp,3} = {\bf e}_{6} .
\end{align}
On the other hand, the sublattice $\Lambda_{-\cR,\inv}$, 
which is invariant under $-\cR$, 
is given by 
\begin{align}
\begin{split}
\Lambda_{-\cR, \inv} &=
\Big\{ \sum_{i=1}^3 n_{\inv,i} {\bf \alpha}_{\inv,i}
\,\Big|\,n_{\inv,i} \in \Z
\Big\}
, \\
{\bf \alpha}_{\inv,1} &= {\bf e}_{2} - {\bf e}_4  , \ \ \ 
{\bf \alpha}_{\inv,2} = {\bf e}_{4} - {\bf e}_6  , \ \ \ 
{\bf \alpha}_{\inv,3} = {\bf e}_4 + {\bf e}_{6} .
\end{split}
\end{align}
Then we can easily evaluate the number of the O6-planes for the {\bf
  CCC} model as
\bea
N_{\text{O6}} =2^3 
\cdot \frac{\vol(\Lambda_{-\cR,\perp})}{\vol(\Lambda_{-\cR,\inv})} =4
\; . 
\eea
In the same way, we consider the {\bf CCD} model.
The lattices $\Lambda_{-\cR,\perp}$ is given by 
\begin{align}
\begin{split}
\Lambda_{-\cR,\perp} &=
\Big\{
\sum_{i=1}^3 n_{\perp,i} {\bf \alpha}_{\perp,i}\,\Big|\,n_{\perp,i} \in \Z \Big\}
, \\
{\bf \alpha}_{\perp,1} &=  {\bf e}_2 , \ \ \ 
{\bf \alpha}_{\perp,2} =  {\bf e}_4 , \ \ \
{\bf \alpha}_{\perp,3} = \half ({\bf e}_5 - {\bf e}_6) ,
\end{split}
\end{align}
and $\Lambda_{-\cR,\inv}$ is given by 
\begin{align}
\begin{split}
\Lambda_{-\cR,\inv} &=
\Big\{ \sum_{i=1}^3 n_{\inv,i} {\bf \alpha}_{\inv,i}
\,\Big|\, n_{\inv,i} \in \Z \Big\}
, \\
{\bf \alpha}_{\inv,1} &= {\bf e}_2 - {\bf e}_4 , \ \ \ 
{\bf \alpha}_{\inv,2} =  {\bf e}_2 + {\bf e}_4 , \ \ \ 
{\bf \alpha}_{\inv,3} =  {\bf e}_5 - {\bf e}_6 . 
\end{split}
\end{align}
Then we obtain $N_{\text{O6}}=2$. 
Substituting these numbers 
into the RR-tadpole cancellation condition (\ref{o6-condition}),
we easily obtain the number of D-branes.
Here we summarize the data of the orientifolds on the non-factorizable
$D_6$ lattice:
\begin{align}
\begin{array}{l@{\;: \ \ }r}
\text{\bf CCC}& (N-16)^2=0 , \\
\text{\bf CCD}& (N-8)^2=0 , \\
\text{\bf CDD}& (N-4)^2=0  , \\
\text{\bf DDD}& (N-8)^2=0 .
\end{array}
\end{align}
These results completely agree with the ones in \cite{Forste:2007zb}.
The gauge group of these models are 
$SO(16)$, $SO(8)$, $SO(4)$ and $SO(8)$, respectively. 
For models on non-factorizable tori, 
the closed string spectra are the same as that of factorizable models. 

We evaluated the the number of O6-planes 
$N_{\text{O6}}$ according to the Lefschetz fixed point theorem, 
and from (\ref{o6-number}) 
this give the necessary and sufficient condition for the RR-tadpole condition. 
This analysis is generic and 
provides quite a simple rule to calculate the number of O-planes and
D-branes in orientifold models on non-factorizable tori in Type II string theory.

\section{Supersymmetric ${\mathbb Z}_N \times {\mathbb Z}_M$
  orientifold models}
\label{ZZorient}

In this section let us consider 
Type IIA supersymmetric orientifold models on orbifolds
and describe the way to deal with orientifolds 
on non-factorizable lattices. 
Since the contributions of the RR-tadpole from untwisted states 
are calculated in the same way as the ones of the orientifolds on tori, 
we can easily count the numbers of D-branes 
via the Lefschetz fixed point theorem (\ref{o6-number}).
We also provide detail calculations of the RR-tadpole cancellation condition 
on ${\mathbb Z}_4 \times {\mathbb Z}_2$ 
and ${\mathbb Z}_2 \times {\mathbb Z}_2$ orbifolds.

\subsection{Orbifolds and orientifolds}

In the previous section 
we showed general expressions for orientifolds on non-factorizable
tori (\ref{orientifold}). 
Here let us consider orientifold models on orbifolds given by
\beq
\frac{\text{Type IIA on $T^6$}}{\Omega \cR \times {\mathbb Z}_N \times
  {\mathbb Z}_M}. \label{orbifold}
\eeq
An orbifold is defined as a quotient of torus 
over a discrete set of isometries of the torus \cite{Dixon:1986jc}, 
called the point group $P$, i.e.,
\beq
\cO = T^6/P={\mathbb R}^6/S. 
\eeq
Here $S$ is called the space group, and is the semi-direct product of the 
point group $P$ and the translation group $T$. 
${\mathbb Z}_N$ orbifolds on the Lie root lattices have been classified 
in terms of the Coxeter elements or the generalized Coxeter elements. 
In the case of ${\mathbb Z}_N \times {\mathbb Z}_M$ orbifolds,
the (generalized) Coxeter elements yield only orbifolds on factorizable lattices. 
Recently, however,
${\mathbb Z}_N \times {\mathbb Z}_M$ orbifolds on non-factorizable lattices  
were investigated in heterotic strings \cite{Takahashi:2007qc}.
We apply their analyses to Type IIA orientifold models.

Since the point group $P$ of orbifold must act crystallographically on
the lattice,
we choose these elements from 
the group generated by the Weyl reflection (\ref{weyl-ref})
and the outer automorphisms $G_{\text{out}}$. 
In the case of the ${\mathbb Z}_N \times {\mathbb Z}_M$ orbifold 
on a Lie root lattice, 
the point group elements of the orbifold can be defined by 
two commutative elements in the group 
generated from Weyl group and the outer automorphisms, i.e,
\beq
[ \theta , \phi ] = 0 , \ls
\theta,~\phi \in \{ \cW,G_{\text{out}} \}.
\eeq
On the complex coordinates of the torus $T^6$, 
the point group elements of the orbifold act in such a way as
\begin{align}
\begin{array}{r@{\; : \ }lcl}
\theta & (z_1,~z_2,~z_3) &\raw& 
(e^{2\pi i v_1} z_1,~e^{2\pi i v_2} z_2,~e^{2\pi i v_3} z_3), \\
\phi & (z_1,~z_2,~z_3) &\raw& 
(e^{2\pi i w_1} z_1,~e^{2\pi i w_2} z_2,~e^{2\pi i w_3} z_3), 
\end{array}
\end{align}
where $(v_1,v_2,v_3)$ and $(w_1,w_2,w_3)$ are twists of an orbifold.  
We consider orientifold models with $\cN =1$ supersymmetry as follows:
The requirement of $SU(3)$ holonomy can be phrased as 
invariance of the $(3,0)$-form $\Omega = dz_1 \wedge dz_2 \wedge dz_3$, 
and leads to 
\beq
v_1 + v_2 + v_3 = w_1 + w_2 + w_3 = 0 . 
\eeq
The twists of the ${\mathbb Z}_N \times {\mathbb Z}_M$ orbifolds 
which are compatible with $\cN=1$ supersymmetric orientifolds 
are listed in Table \ref{zmzn-twist}.
%
%

\begin{table}[htb]
\renewcommand{\arraystretch}{1.5}
\normalsize
\begin{center}
\begin{tabular}{|c|c|c||c|c|c|}
\hline
 & $(v_1,v_2,v_3)$ & $(w_1,w_2,w_3)$ & & $(v_1,v_2,v_3)$ & $(w_1,w_2,w_3)$ \\
\hline
\hline
${\mathbb Z}_2 \times {\mathbb Z}_2$ 
& $(\frac12, -\frac12, 0)$ & $(0, \frac12, -\frac12)$ &
${\mathbb Z}_2 \times {\mathbb Z}_4$ 
& $(\frac12, -\frac12, 0)$ & $(0, \frac14, -\frac14)$ \\
${\mathbb Z}_2 \times {\mathbb Z}_6$ 
& $(\frac12, -\frac12, 0)$ & $(0, \frac16, -\frac16)$ &
${\mathbb Z}_2 \times {\mathbb Z}_6^\prime$ 
& $(\frac12, -\frac12, 0)$ & $(\frac16, -\frac13, \frac16)$ \\
${\mathbb Z}_3 \times {\mathbb Z}_3$ 
& $(\frac13, -\frac13, 0)$ & $(0, \frac13, -\frac13)$ &
${\mathbb Z}_3 \times {\mathbb Z}_6$ 
& $(\frac13, -\frac13, 0)$ & $(0, \frac16, -\frac16)$ \\
${\mathbb Z}_4 \times {\mathbb Z}_4$ 
& $(\frac14, -\frac14, 0)$ & $(0, \frac14, -\frac14)$ &
${\mathbb Z}_6 \times {\mathbb Z}_6$ 
& $(\frac16, -\frac16, 0)$ & $(0, \frac16, -\frac16)$ \\
\hline
\end{tabular}
\end{center}
\caption{\it Twists of ${\mathbb Z}_N \times {\mathbb Z}_M$ orbifolds.}
\label{zmzn-twist}
\end{table}

\noi
As explained in Appendix \ref{App-LRL} 
there are twelve distinct classes of 
non-factorizable lattices, see Table \ref{all-lattices}.
The ${\mathbb Z}_2 \times {\mathbb Z}_2$, ${\mathbb Z}_4 \times {\mathbb Z}_2$ and 
${\mathbb Z}_4 \times {\mathbb Z}_4$ orbifolds are allowed 
on these non-factorizable lattices 
(see Table \ref{zmzn-orbifold} in appendix \ref{App-LRL}). 
The series of generators $\theta$ and $\phi$ of the
${\mathbb Z}_N \times {\mathbb Z}_M$ orbifold as well as 
the action $\OR$ consist of the orientifold group: 
\beq
\Big\{ \theta^{k_1} \phi^{k_2},~ \OR \theta^{k_1} \phi^{k_2} \,\Big|\, 
{k_1}=0,\dots,N;~{k_2}=0,\dots,M \Big\}, 
\eeq
These elements appear 
in the following string one-loop amplitudes as insertions \cite{Forste:2000hx}, 
\bsubeq
\begin{align}
{\cal K} &= 4c\int_0^\infty \frac{dt}{t^3} 
\Tr_{\text{closed}}
\left(
  \frac{\Omega {\cal R}}{2}\mbox{\bf P}\left(\frac{1
  +\left(-1\right)^{F}}{2}\right)\left(-1\right)^{\text{\bf S}} 
e^{-2\pi t\left( L_0+\bar{L}_0\right)}\right)
  , \label{kb} \\
{\cal A} &= c \int_0^\infty
  \frac{dt}{t^3} \Tr_{\text{open}}
\left( \frac{1}{2}
  \mbox{\bf P}\left(\frac{1 
  +\left(-1\right)^{F}}{2}\right)\left(-1\right)^{\text{\bf S}} 
e^{-2\pi t L_0}\right) 
, \label{an} \\
{\cal M} &=  c \int_0^\infty
  \frac{dt}{t^3} 
\Tr_{\text{open}}\left(\frac{\Omega {\cal R}}{2}\mbox{\bf P}\left(\frac{1 
  +\left(-1\right)^{F}}{2}\right)\left(-1\right)^{\text{\bf S}} 
e^{-2\pi t L_0}\right)
. \label{ms}
\end{align}
\esubeq
Here
\begin{equation}
\mbox{\bf P} = \left(\frac{1+ \theta + \cdots + \theta^{N-1}}{N}\right) 
\left(\frac{1+ \phi + \cdots + \phi^{M-1}}{M}\right)
.
\end{equation}
After extracting the RR-tadpoles, 
the insertion of $\OR\theta^{k_1} \phi^{k_2}$ in the Klein bottle amplitude 
corresponds to the contribution from O-planes fixed by $\cR\theta^{k_1} \phi^{k_2}$.
Since in the $\OR\theta^{k_1} \phi^{k_2}$ insertion 
the contributions from untwisted sectors are 
calculated in the same way as the cases of tori in section \ref{ONFT}, 
we obtain the necessary condition (\ref{o6-condition}) 
for the RR-tadpole cancellation by D-branes parallel to the O-planes. 
From this necessary condition, 
we obtain all the numbers of O-planes and D-branes on the orbifold. 
In the next subsection we will demonstrate a few examples of 
${\mathbb Z}_4 \times {\mathbb Z}_2$ orientifold models, 
and evaluate the RR-tadpole cancellation condition.

\subsection{${\mathbb Z}_4 \times {\mathbb Z}_2$ model}

Here we discuss the ${\mathbb Z}_4 \times {\mathbb Z}_2$ orientifold 
model on the Lie root lattice $D_6$ (\ref{d6-lattice}) 
in detail because in this case all possible subtleties show up. 

There exists only one distinct ${\mathbb Z}_4 \times {\mathbb Z}_2$ orbifold 
on $D_6$, whose point group elements $\theta$ and $\phi$ are given by
\bsubeq
\begin{align}
\theta:& \ \ 
\left\{ \begin{array}{rcll}
{\bf e}_1 &\raw& {\bf e}_2 &\raw -{\bf e}_1 \\
{\bf e}_3 &\raw& -{\bf e}_4 &\raw -{\bf e}_3  \\
{\bf e}_{5} &\raw& {\bf e}_{5} &  \\
{\bf e}_{6} &\raw& {\bf e}_{6} & 
\end{array} \right. 
\LS
\phi: \ \ 
\left\{ \begin{array}{rcl}
 {\bf e}_1 &\raw& {\bf e}_1  \\
{\bf e}_2 &\raw& {\bf e}_2  \\
{\bf e}_i &\raw& -{\bf e}_i \ls i=3,4,5,6 
\end{array} \right. 
\end{align}
or, in matrix representation, by 
\beq
\theta :~
\left( \begin{array}{cccccc}
   0 & -1 &  0 &  0 &  0 &  0 \\
   1 &  0 & 0 &  0 &  0 &  0 \\
   0 &  0 &  0 & 1 &  0 &  0 \\
   0 &  0 &  -1 &  0 & 0 &  0 \\
   0 &  0 &  0 &  0 &  1 & 0 \\
   0 &  0 &  0 &  0 &  0 &  1 \\
\end{array} \right) ,
~~~~~
\phi :~
\left( \begin{array}{cccccc}
   1 & 0 &  0 &  0 &  0 &  0 \\
   0 &  1 & 0 &  0 &  0 &  0 \\
   0 &  0 &  -1 & 0 &  0 &  0 \\
   0 &  0 &  0 &  -1 & 0 &  0 \\
   0 &  0 &  0 &  0 &  -1 & 0 \\
   0 &  0 &  0 &  0 &  0 &  -1 \\
\end{array} \right) .
\label{z4z2}
\eeq
\esubeq
By using the above elements
we can show all the orientifold actions which preserve $\cN=1$ supersymmetry 
by means of {\bf C} and {\bf D} actions. 
For example, the reflection $\cR$ on the {\bf DDC} model is given by 
\beq
\cR :~
\left( \begin{array}{cccccc}
   0 & 1 &  0 &  0 &  0 &  0 \\
   1 &  0 & 0 &  0 &  0 &  0 \\
   0 &  0 &  0 & 1 &  0 &  0 \\
   0 &  0 &  1 &  0 & 0 &  0 \\
   0 &  0 &  0 &  0 &  1 & 0 \\
   0 &  0 &  0 &  0 &  0 &  -1 \\
\end{array} \right) 
\equiv (\bb,\bb,\ba) ,
\label{DDC}
\eeq
where we used an abbreviation defined by 
\begin{eqnarray}
 (\mathbf{m}_{1}, \mathbf{m}_{2}, \mathbf{m}_{3} ) \equiv
\left( \begin{array}{ccc}
  \mathbf{m}_{1} & \mathbf{0} & \mathbf{0}\\
  \mathbf{0} & \mathbf{m}_{2} & \mathbf{0}\\
  \mathbf{0} & \mathbf{0} & \mathbf{m}_{3}
 \end{array} \right) ~~
 \text{with $\mathbf{m}_{i} \in \{\pm \ba , \pm \bb , \pm \b1 \}$}
\end{eqnarray}
and 
\bea 
 \mathbf{a} \equiv 
\left( \begin{array}{cc}
  1 & 0\\
  0 &-1
\end{array} \right) , \ \ 
 \mathbf{b} \equiv
\left( \begin{array}{cc}
  0 & 1\\
  1 & 0
\end{array} \right) , \ \ 
 \mathbf{1} \equiv
\left( \begin{array}{cc}
  1 & 0\\
  0 & 1
\end{array} \right) , \ \ 
 \mathbf{0} \equiv
\left( \begin{array}{cc}
  0 & 0\\
  0 & 0
\end{array} \right) .
\label{ab10}
\eea
From the Lefschetz fixed point theorem (\ref{o6-number}), 
the number of O6-plane fixed by $\cR$ is given as
$N_{\text{O6}} =1$. 
If we put four D-branes parallel to this $\cR$-fixed
O6-plane, the RR-tadpole of this model will be cancelled. 
Similarly,
the element $\cR \theta=(\ba,-\ba,\ba)$ gives $N_{\text{O6}} = 4$,
whose tadpole is cancelled by 
sixteen D-branes parallel to this four $\cR \theta$-fixed O6-planes. 
We similarly evaluate the cases for the other elements of the orientifold group. 
The relations between the orientifold group elements and 
the numbers of O-planes are summarized in Table \ref{o-action}. 
\begin{table}[htb]
\renewcommand{\arraystretch}{1.3}
\normalsize
\begin{center}
\begin{tabular}{|l|c|}
\hline
Orientifold elements of $\cR$ & \# of O6-planes \\
\hline
\hline
$(\pm \ba,\pm \ba,\pm \ba)$, 
$(\underline{\b1 , -\b1 ,\pm \ba})$ & 4 \\
\hline
$(\underline{\pm \ba ,\pm \ba ,\pm \bb})$, 
$(\underline{\b1 , -\b1 ,\pm \bb})$ & \multirow{2}{*}{2} \\
$(\pm \bb,\pm \bb,\pm \bb)$ &   \\
\hline
$(\underline{\pm \ba ,\pm \bb ,\pm \bb})$ & 1 \\
\hline
\end{tabular}
\end{center}
\caption{\it Orientifold group elements and the numbers of O6-planes 
on the $D_6$ lattice. 
The underline indicates a symmetry under the cyclic permutation.}
\label{o-action}
\end{table}

\noi
Since the $\Z_4$ action changes the directions of the O-planes 
by angle of $\theta^{1/2}$ in the following way:
\beq
\cR \theta = \theta^{-1/2} \cR \theta^{1/2}.
\label{half-rotate}
\eeq
This action generates the exchange between 
the action {\bf C} and {\bf D} each other. 
Then we can see that {\bf CCC} and {\bf DDC}, {\bf CCD} and {\bf DDD}, 
{\bf CDD} and {\bf DCD} models are equivalent with each other, respectively. 
In the case of the $\mathbf{CCC}$ model, for example, 
two different numbers of O6-planes
 appear since the orientifold group elements in $\cR$,
 $\cR\theta^2$, $\cR \phi$ and $\cR\theta^2\phi$ are given by $(\pm \ba,
 \pm \ba, \pm \ba)$,  whereas the elements in
 $\cR \theta$, $\cR \theta^3$, $\cR \theta \phi$ and $\cR \theta^3
 \phi$ are given by $(\pm \bb, \pm \bb, \pm \ba)$.
Analyzing such actions, 
we obtain all the models for ${\mathbb Z}_4 \times {\mathbb Z}_2$ orientifolds 
on $D_6$ lattice, listed in Table \ref{z4z2d6}. 
\begin{table}[htb]
\renewcommand{\arraystretch}{1.3}
\normalsize
\begin{center}
\begin{tabular}{|c|c|c|c|c|}
 \hline
 \multirow{2}{*}{Lattice}
 & \multirow{2}{*}{Label}
 & \multirow{2}{*}{reps.\,of $\cR$}
 & \multicolumn{2}{|c|}{\# of O6-planes}\\
 \cline{4-5}
 &&& \ $\cR$, $\cR\theta^2$, $\cR\phi$, $\cR\theta^2\phi$
 & $\cR\theta$, $\cR\theta^3$, $\cR\theta\phi$, $\cR\theta^3\phi$  \\
 \hline
 \hline
 \multirow{4}{*}{$D_6$}
 & $\mathbf{CCC}$
     & $( \mathbf{a},  \mathbf{a},  \mathbf{a})$
	 & $4$ & $1$ \\
 \cline{2-5}
 & $\mathbf{CCD}$
     & $( \mathbf{a},  \mathbf{a},  \mathbf{b})$
	 & $2$ & $2$ \\
 \cline{2-5}
 & $\mathbf{CDD}$
     & $( \mathbf{a},  \mathbf{b},  \mathbf{b})$
	 & $2$ & $2$ \\
 \cline{2-5}
 & $\mathbf{DCC}$
     & $( \mathbf{b},  \mathbf{a},  \mathbf{a})$
	 & $2$ & $2$ \\
 \hline
\end{tabular}
\end{center}
\caption{\it All the $\Z_4 \times \Z_2$ orientifold models on the $D_6$
 lattice.} 
\label{z4z2d6}
\end{table}

We estimated the RR-tadpole cancellation 
by counting the O-planes from the equation (\ref{o6-condition}), 
which is the necessary condition in the case of the orbifold model. 
However it is expected that the RR-tadpoles are cancelled 
even in the orbifold model. 
These countings also give correct results 
for well-investigated 
non-factorizable models on $\Z_N$ orbifolds in \cite{Blumenhagen:2004di} 
and $\Z_2 \times \Z_2$ orbifolds in \cite{Forste:2007zb}. 
We give the explicit results of the 
RR-tadpole cancellation for a few models in the following.

\subsubsection{Klein bottle amplitude} 
\label{42Klein}

First let us evaluate the Klein bottle amplitude of $\Z_4 \times \Z_2$ orientifold 
model on the $D_6$ lattice (\ref{d6-lattice}) 
with the orientifold action  
\beq
\cR = (\bb,\ba,\ba),
\eeq
which gives  the {\bf DCC} model. 
The contribution of the oscillator modes
are equal in any insertions of the orientifold group 
because they act as the unit operator in (\ref{kb}). 
In the $\theta^{n_1} \phi^{n_2}$-twisted sector, 
the oscillator contribution 
is given by $\cK^{(n_1,n_2)}\equiv \cK^{(n_1,k_1)(n_2,k_2)}$ 
(see, for the notation, \cite{Forste:2000hx}). 
We also need the multiplicities $\chi_{\cK}^{(n_1,k_1)(n_2,k_2)}$ 
of the $\theta^{n_1} \phi^{n_2}$-twisted fixed sectors, 
which are invariant under the insertion $\OR \theta^{n_1} \phi^{n_2}$, 
which can be seen in Table \ref{z4z2-fix-k}. 
\begin{table}[htb]
\renewcommand{\arraystretch}{1.3}
\normalsize
\begin{center}
\begin{tabular}{|l|c|c|c|c|}
\hline
multiplicities $\chi_\cK^{(n_1, k_1)(n_2, k_2)}$ 
& {\bf CCC} & {\bf CCD}& {\bf CDD} & {\bf DCC} \\
\hline
\hline
$(0,k_1)(0,k_2)$ & 1 & 1 & 1 & 1 \\
$(2n_1+1,k_1)(0,k_2)$ & 2 & 2  & 2 & 2 \\
$(2n_1,2k_1+1)(0,k_2)$ & 4 & 4  & 4 & 4 \\
$(2n_1,2k_1)(0,k_2)$ & 8 & 8  & 4 & 4 \\
$(0,2k_1+1)(1,k_2)$ & 4 & 4  & 4 & 4 \\
$(0,2k_1)(1,k_2)$ & 8 & 4  & 4 & 4 \\
$(2n_1+1,k_1)(1,k_2)$ & 8 & 8  & 4 & 8 \\
$(2n_1,2k_1+1)(1,k_2)$ & 4 & 4  & 4 & 4 \\
$(2n_1,2k_1)(1,k_2)$ & 8 & 4 & 4 & 4 \\
\hline
\end{tabular}
\end{center}
\caption{\it Multiplicities of the fixed points for the {\bf DCC} and {\bf CCC} models.}
\label{z4z2-fix-k}
\end{table}

\noi
When an action $\theta^{n_1} \phi^{n_2}$ 
does not fix certain directions in the compact space,
the Kaluza-Klein momentum modes and the winding modes appear as the zero modes in
the $\theta^{n_1} \phi^{n_2}$-fixed sector. 
Let us evaluate such zero modes in the $\theta$-twisted sector. 
The $\theta$ invariant sublattice 
$\Lambda^\theta$ is 
expanded in terms of the basis
\beq
\big\{ {\bf e}_5+{\bf e}_6, \ {\bf e}_5-{\bf e}_6 
\big\}.
\eeq
We can see that the $\cR$ invariant dual sublattice 
$(\Lambda^{\theta})^*_{\cR, \inv}$, whose basis is given by 
$\{ 2 {\bf e}_5 \}$, 
and the $-\cR$ invariant sublattice 
$(\Lambda^{\theta})_{-\cR, \inv}$, with its basis $\{2 {\bf e}_6\}$, 
yield the momentum modes and the winding modes in this sector, respectively.
However there are two subtleties in this evaluation, one of which is
caused by the momentum doubling, and the other from the
appearance of the half winding states \cite{Blumenhagen:2004di}.

The former subtlety is caused by the shifts associated to 
the $\OR \theta^{k_1}\phi^{k_2}$ insertions.
In the $\theta$-twisted sector we have two fixed tori given by
\beq
x{\bf e}_5, \ \frac12 ({\bf e}_1+{\bf e}_2+{\bf e}_3+{\bf e}_4) +x {\bf e}_5,
\eeq 
where $x \in \bR$ is a coordinate on the fixed tori. 
Note that the invariance of fixed points or fixed tori 
under $\OR \theta^{k_1}\phi^{k_2}$
is defined modulo the translation generated by the lattice $\Lambda$. 
The $\cR$ insertion acts on the two fixed tori in such a way as 
\begin{align}
\cR: 
\left\{
\begin{array}{rcl}
x {\bf e}_5 &\raw& x {\bf e}_5, \\
\frac12 ({\bf e}_1+{\bf e}_2+{\bf e}_3+{\bf e}_4) +x {\bf e}_5  
&\raw& 
\begin{array}{l}
\frac12 ({\bf e}_1+{\bf e}_2+{\bf e}_3-{\bf e}_4) +x {\bf e}_5  \\ 
\ \  = \frac12 ({\bf e}_1+{\bf e}_2+{\bf e}_3+{\bf e}_4) +(-1+x) {\bf e}_5.
\end{array}
\end{array} \right. 
\label{no-double}
\end{align}
In the latter case, the translation of a lattice shift 
${\bf \alpha}_4={\bf e}_4-{\bf e}_5$  
is accompanied. 
Because a momentum mode $\ket{{\bf p}}$ picks up a phase factor 
$e^{2\pi {\bf p}\cdot l}$ 
under the translation by $l$, 
we generally need phase factors in the amplitudes.
In the case of (\ref{no-double}), 
the phase factor is  ${\bf p} \cdot l = 2 {\bf e}_5 \cdot {\bf e}_5 =2$, 
and does not affect the amplitudes.
If the phase factor is given as $-1$, 
the momentum modes are effectively doubled 
by interference between modes with and without shifts: 
\beq
\sum_n (-1)^n \exp(-\pi t n^2 {\bf p}^2) +\sum_n \exp(-\pi t n^2 {\bf p}^2) 
=2 \sum_n \exp(-4\pi t n^2 {\bf p}^2).
\eeq

The latter subtlety occurs in the winding modes. 
There are special points with the following property: 
\beq
\theta:~ \frac12 ({\bf e}_1 + {\bf e}_2) \raw 
\frac12 ( -{\bf e}_1 + {\bf e}_2) = \frac12 ( {\bf e}_1 + {\bf e}_2) +{\bf e}_6, 
\eeq
where we used a lattice shift given by ${\bf e}_1+{\bf e}_6$. 
The point does not lie on the $\theta$-fixed tori, 
whereas this shift {\it does} generate the winding modes: 
\beq
X(\sigma,\tau) = \frac12 ({\bf e}_1 + {\bf e}_2) + \frac{\sigma}{2\pi} {\bf e}_6 
+ (\text{$\tau$ dependence}),
\eeq
There are two points $\frac12 ({\bf e}_1 \pm {\bf e}_2)$ 
which are invariant under the action $\cR$, 
and the multiplicity is equal to that of the $\theta$-fixed tori 
which are also invariant under $\cR$. 

Therefore we conclude that 
the zero modes in the $\theta$-twisted sector with $\cR$ insertion are
given by the following vectors:
\bea
{\bf p} = 2n {\bf e}_5 , \ls
{\bf w} = m {\bf e}_6 , 
\eea
where $n,m \in \bZ$. 
In the notation of (\ref{zero-loop}), 
the zero mode contributions in the Klein bottle amplitude is $\cL_{2,\frac12}$.
In a similar way we can evaluate the other twisted sectors in the
orbifold model. 
Note that for non-factorizable orbifolds 
the zero mode contributions depend on the 
insertion $\OR\theta^{k_1}\phi^{k_2}$. 
In the $\phi$-twisted sector we have $\cL_{2,\frac12}$ 
for the $\OR\theta^{2k_1}\phi^{k_2}$ insertions, 
and $\cL_{4,1}$ for the $\OR\theta^{2k_1+1}\phi^{k_2}$ insertions.

Next we evaluate the zero mode contribution 
from the untwisted sector given in (\ref{MK}). 
The basis of dual lattice ${\bf \alpha}^* \in \Lambda^*$, 
which is defined 
by ${\bf \alpha}^*_i \cdot {\bf \alpha}_j =\delta_{ij}$, 
is given as
\bea
\begin{array}{rcl}
{\bf \alpha}^*_1 &=& {\bf e}_1, \\
{\bf \alpha}^*_2 &=& {\bf e}_1 + {\bf e}_2, \\
{\bf \alpha}^*_3 &=& {\bf e}_1 + {\bf e}_2 + {\bf e}_3, \\
{\bf \alpha}^*_4 &=& {\bf e}_1 + {\bf e}_2 + {\bf e}_3 + {\bf e}_4, \\
{\bf \alpha}^*_5 &=& \frac12 ({\bf e}_1  + {\bf e}_2 + {\bf e}_3 + {\bf e}_4 + {\bf e}_5 -{\bf e}_6), \\
{\bf \alpha}^*_6 &=& \frac12 ({\bf e}_1  + {\bf e}_2 + {\bf e}_3 +
{\bf e}_4 + {\bf e}_5 +{\bf e}_6) .
\end{array}
\eea
Then the $\cR$ invariant dual sublattice 
$\Lambda^*_{\cR,\inv}$ in (\ref{MK}), 
which yields the momentum modes in the Kaluza-Klein states,
is expanded by the basis
\beq
\big \{ {\bf e}_{1}+{\bf e}_{2}, \ {\bf e}_{3}, \ {\bf e}_{5}
\big\}
.
\eeq
In the same way, the $-\cR$ invariant lattice 
$\Lambda_{-\cR,\inv}$ in (\ref{WK}) yielding 
the winding states is expanded by
\beq
\big\{ {\bf e}_{1}-{\bf e}_{2}, \ 2 {\bf e}_{4}, \  2 {\bf e}_{6}
\big\}
.
\eeq
Substituting these elements into (\ref{zero-part}),
we obtain the zero mode contribution $\cL_\cK^{(0,0)}$. 
Its modular transformation is given by the factors 
\begin{align}
\sqrt{\det M^{\cK}} &= \vol(\Lambda^*_{\cR,\inv})=\sqrt{2}, \\
\sqrt{\det W^{\cK}} &= \vol(\Lambda_{-\cR,\inv})=4\sqrt{2}. \nn
\label{k-mw}
\end{align}
We also need the zero mode contributions 
with the other insertions $\OR\theta^{k_1}\phi^{k_2}$. 
Since these elements are given by 
$\cR\theta^{k_1}\phi^{k_2}= ( \pm \mathbf{a}, \pm \mathbf{b}, \pm \mathbf{b})$ 
for the {\bf DCC} model,
we have the same results as that of the $\cR$ insertion.

We obtained all the ingredients to write down 
the Klein bottle amplitude for the {\bf DCC} model, which are
summarized as 
\begin{align}
    {\cal K} = c(1_{\text{RR}}-1_{\text{NSNS}}) 
    &\int_{0}^{\infty} 
    \frac{dt}{t^3} 
     \Big( \cL_\cK^{(0,0)} \cK^{(0,0)} 
    + 2\cL_{2,\frac12}\cK^{(1,0)} 
    + 4\cL_{2,\frac12}\cK^{(2,0)} 
    + 2\cL_{2\frac12}\cK^{(3,0)} \nn\\
    &\LS
    + \half \big( 8 \cL_{4,1}+4 \cL_{2,\frac12} \big) \cK^{(0,1)} 
    + 8\cK^{(1,1)} \nn\\
    &\LS + \half \big( 8 \cL_{4,1}+4 \cL_{2,\frac12} \big) \cK^{(2,1)} 
    + 8\cK^{(3,1)} \Big).
\end{align}
Its modular transformation to the tree channel is 
\begin{align}
    \ctK = 16c(1_{\text{RR}}-1_{\text{NSNS}})
    &\int_{0}^{\infty} dl 
    \Big( \ctL_\cK^{(0,0)} \ctK^{(0,0)} 
    - 2\ctL_{2,8}\ctK^{(1,0)} 
    - 4\ctL_{2,8}\ctK^{(2,0)} 
    - 2\ctL_{2,8}\ctK^{(3,0)} \nn\\
    &\LS - 2(\ctL_{1,4}+\ctL_{2,8})\cK^{(0,1)} 
    + 4\ctK^{(1,1)} \nn\\
    &\LS - 2(\ctL_{1,4}+\ctL_{2,8})\ctK^{(2,1)} 
    - 4\ctK^{(3,1)} \Big). 
\label{Z4Z2-ctK}
\end{align}
Note that in the IR limit $l \to \infty$,
the zero mode contributions 
$\ctL_{\cal K}^{(0,0)}$ and $\ctL_{\alpha, \beta}$ 
in the tree channel (\ref{Z4Z2-ctK})
go to unity, then we obtain $2(\ctL_{1,4}+\ctL_{2,8}) \raw 4$. 
Then we observe that the prefactors are given 
by the complete projector \cite{Blumenhagen:1999md}
\beq
  \prod_{i=1,~n_1v_i + n_2 w_i \neq 0}^3 \big
  ( -2\sin(\pi n_1v_i + \pi n_2w_i) \big).
  \label{complete}
\eeq
This relation implies that only the untwisted sector contributes to
the RR-tadpole.

\subsubsection{Annulus amplitude}

In order to cancel the RR-tadpole 
we introduce D-branes parallel to O-planes. 
We attach a label $(i_1, i_2)$ to 
a stock of D-branes  
which is invariant under the orientifold action $\cR\theta^{i_1}\phi^{i_2}$,
and define 
that $(0,0)$ denotes D-branes invariant under the action $\cR$. 
The three-cycle wrapped by the brane $(0,0)$ is given by the 
$\cR$ invariant lattice $\Lambda_{\cR,\inv}$ whose basis is given by
\beq
\big\{ {\bf e}_1+{\bf e}_2, \ {\bf e}_3-{\bf e}_5, \ 
{\bf e}_3+{\bf e}_5 
\big\} .
\eeq
From (\ref{half-rotate}) 
the brane $(1,0)$ is rotated by half the angle of $\theta$ 
with respect to the brane $(0,0)$.  
The three-cycle wrapped by the brane $(1,0)$ is given by the 
$\cR \theta$ invariant lattice $\Lambda_{\cR\theta,\inv}$ whose basis is given by
\beq
\big\{
{\bf e}_1-{\bf e}_5, \ {\bf e}_3+{\bf e}_4, \ {\bf e}_1+{\bf e}_5 
\big\} .
\eeq
An open string stretching 
from brane $(i_1,i_2)$ to brane $(i_1-n_1,i_2-n_2)$ 
is localized at intersection of D-branes.
It is convenient to call such a state the $\theta^{n_1}\phi^{n_2}$-twisted sector. 

The three-cycles of brane $(0,0)$ and brane $(1,0)$ share
a common direction, 
and the lattice vector in this direction is given by $2{\bf e}_5$.
The momentum modes are obtained from the dual of the vector $2{\bf e}_5$
in such a way as
\beq
{\bf p} = \frac{n}{\sqrt{2}} {\bf e}_5.
\eeq
where $n \in \bZ$. 
The basis of the winding modes is related to the distances of 
the parallel D-branes. 
Because we put D-branes parallel to the O-planes, 
the shortest distance corresponds 
to the lattice vector projected by the actions $-\cR$ and $-\cR\theta$, 
i.e., $\Lambda_{\cR,\perp} \cap \Lambda_{\cR\theta,\perp}$.
Then the winding modes are  
\beq
{\bf w} = \frac{n}{\sqrt{2}} {\bf e}_6.
\eeq
Then zero mode contribution of the $\theta$-twisted sector is 
expressed as $\cL_{1,1}$. 

Let us explain one more case of the $\phi$-twisted sector. 
The winding modes are given as ``half winding-like'' modes, 
and are also given by the projected lattice 
$\Lambda_{\cR,\perp} \cap \Lambda_{\cR\phi,\perp}$ whose basis is
\beq 
\Big\{ 
\frac{1}{2\sqrt{2}} ({\bf e}_1+{\bf e}_2) 
\Big\}
.
\eeq
Then the zero modes of open string stretching between
the brane $(0,0)$ and the brane $(0,1)$ are given by
\bea
{\bf p} = \frac{n}{\sqrt{2}} ({\bf e}_1+{\bf e}_2), \ls
{\bf w} = \frac{n}{2\sqrt{2}} ({\bf e}_1+{\bf e}_2).
\eea
The zero mode contribution in the annulus amplitude is 
expressed as $\cL_{2,\frac12}$. 
The other zero modes are calculated in a similar way.

Since in the $\theta^{n_1}\phi^{n_2}$-twisted sector 
the contributions from the oscillator modes do not depend on branes $(i_1,i_2)$, 
they are given as $\cA^{(n_1,k_1)(n_2,k_2)}$. 
The insertions of $\id$, $\theta^2$, $\phi$ and $\theta^2\phi$  
leave D-branes invariant, 
and perform non-trivial actions on the Chan-Paton factors 
described as $\gamma^{(i_1,i_2)}_{k_1,k_2}$, which appear 
in the amplitude as
\beq
\tr \Big( \gamma^{(i_1-n_1,i_2-n_2)}_{k_1,k_2} \Big)
\tr \Big( \gamma^{(i_1,i_2)}_{k_1,k_2} \Big)^{-1}
\eeq  
in the $\theta^{n_1}\phi^{n_2}$-twisted sector.  
Sectors of $k_1 \not=0$ or $k_2 \not=0$
cannot be cancelled by the other diagrams. 
Therefore the $\Z_2$ twisted tadpole cancellation condition 
is required \cite{Gimon:1996rq,Forste:2000hx}: 
\beq
\tr \left( \gamma^{(i_1,i_2)}_{2,0} \right)=
\tr \left( \gamma^{(i_1,i_2)}_{0,1} \right)=
\tr \left( \gamma^{(i_1,i_2)}_{2,1} \right)=0 .
\label{z2-condition}
\eeq

We should also evaluate the multiplicities $\chi_{\cM}$ of the open string states, 
which are given by the intersection number of D-branes. 
The intersection numbers of two branes can be 
obtained by the determinant of vectors $v_i$ and $v_i'$ 
giving the three-cycles in respective D-branes \cite{Blumenhagen:2004di}.
These vectors can be expanded in terms of the lattice basis as 
$v_i = \sum v_{ij} {\bf \alpha}_j$. 
Then the intersection number is 
\beq
I= \det \left(
\begin{array}{cccc}
v_{11} & v_{12} & \cdots & v_{16} \\
v_{21} & v_{22} & \cdots & v_{26} \\
\vdots &   &   & \vdots \\
v_{31}^\prime & v_{32}^\prime & \cdots & v_{36}^\prime 
\end{array} \right) .
\eeq
Owing to the above $\Z_2$ twisted tadpole condition, 
it is sufficient to consider 
the intersection number $\chi_{\cM}$ for $k_1=k_2=0$,
which are given in Table \ref{z4z2-fix-a}. 
\begin{table}[htb]
\renewcommand{\arraystretch}{1.3}
\normalsize
\begin{center}
\begin{tabular}{|l|c|c|c|c|}
\hline
$\chi_\cA$ & {\bf CCC} & {\bf CCD}& {\bf CDD} & {\bf DCC} \\
\hline
\hline
($i_1$,$i_2$)--($i_1$,$i_2$) & 1 & 1  & 1 & 1  \\
($i_1$,$i_2$)--($i_1$+1,$i_2$) & 1 & 1  & 2 & 1  \\
($2i_1$+1,$i_2$)--($2i_1$+3,$i_2$) & 4 & 2  & 4 & 2 \\
($2i_1$,$i_2$)--($2i_1$+2,$i_2$) & 1 & 2  & 4 & 2 \\
($2i_1$+1,$i_2$)--($2i_1$+1,$i_2$+1) & 4 & 2  & 4 & 2 \\
($2i_1$,$i_2$)--($2i_1$,$i_2$+1) & 1 & 2  & 4 & 2 \\
($i_1$,$i_2$)--($i_1$+1,$i_2$+1) & 2 & 2  & 4 & 2  \\
($2i_1$+1,$i_2$)--($2i_1$+3,$i_2$+1) & 4 & 2 & 4 & 2 \\
($2i_1$,$i_2$)--($2i_1$+2,$i_2$+1) & 1 & 2 & 4 & 2 \\
\hline
\end{tabular}
\end{center}
\caption{\it Intersection numbers in the annulus amplitudes in the
  {\bf DCC} and {\bf CCC} models.}
\label{z4z2-fix-a}
\end{table}

The contribution from the zero modes of the untwisted sector 
is obtained from (\ref{MA}) and (\ref{WA}). 
For the brane $(0,0)$, which is paralell to the $\cR$-fixed O6-plane
,it is
\begin{align}
\sqrt{\det M^{\cA}} &= \vol(\Lambda^*_{\cR,\perp})=4, \\
\sqrt{\det W^{\cA}} &= \vol(\Lambda_{-\cR,\perp})=4. \nn
\label{a-mw}
\end{align}
The contributions from the other branes $(i_1,i_2)$ give the same values. 
These values appear 
in prefactors of the amplitude after the modular transformation.

Summarizing the above,
we obtain the annulus amplitude for the {\bf DCC} model 
\begin{align}
    {\cal A} = \frac{N^2 c}{4} (1_{\text{RR}}-1_{\text{NSNS}}) 
    &\int_{0}^{\infty} 
    \frac{dt}{t^3} 
     \Big( \cL_\cA^{(0,0)} \cA^{(0,0)} 
    + \cL_{1,1}\cA^{(1,0)} 
    + 2\cL_{1,1}\cA^{(2,0)} 
    + \cL_{1,1}\cA^{(3,0)} \nn \\
    &\LS + (\cL_{2,\frac12}+\cL_{1,1})\cK^{(0,1)} 
    + 2\cA^{(1,1)} \nn \\ 
    &\LS + (\cL_{1,1}+\cL_{2,\frac12})\cA^{(2,1)} 
    + 2\cA^{(3,1)} \Big).
\end{align}
where $\cA^{(n_1,n_2)} \equiv \cA^{(n_1,0)(n_2,0)}$. 
The modular transformation to the amplitude in the tree channel yields 
\begin{align}
    \ctA = \frac{N^2 c}{4} (1_{\text{RR}}-1_{\text{NSNS}}) 
    &\int_{0}^{\infty} dl 
     \Big( \ctL_\cA^{(0,0)} \ctA^{(0,0)} 
    - 2\ctL_{2,2}\ctA^{(1,0)} 
    - 4\ctL_{2,2}\ctA^{(2,0)} 
    - 2\ctL_{2,2}\ctA^{(3,0)} \nn\\
    &\LS - 2(\ctL_{1,4}+\ctL_{2,2})\cA^{(0,1)} 
    + 4\ctA^{(1,1)} \nn\\
    &\LS - 2(\ctL_{2,2}+\ctL_{1,4})\ctA^{(2,1)} 
    - 4\ctA^{(3,1)} \Big).
\label{Z4Z2-ctA}
\end{align}
We again observe the complete projector in the IR limit.

\subsubsection{M\"{o}bius strip amplitude}

The amplitude of the M\"obius strip (\ref{ms})
includes the insertion of $\OR$,
and string states should be invariant under these orientifold actions.
In $\theta^{n_1}\phi^{n_2}$-twisted sector,
the insertion $\OR\theta^{k_1}\phi^{k_2}$ acts on
open strings stretching from brane $(i_1,i_2)$ to brane $(i_1-n_1,i_2-n_2)$ as
\beq
\OR\theta^{k_1}\phi^{k_2}:~
[(i_1,i_2)(i_1-n_1,i_2-n_2)] \raw
[(-i_1+n_1-2k_1,-i_2+n_2-2k_2)(-i_1-2k_1,-i_2-2k_2)],
\eeq
Therefore in the $\Z_4 \times \Z_2$ orbifold case the following
conditions are required:
\bsubeq \label{m-condition} 
\begin{align}
2(i_1 +k_1)-n_1 &= 0 \pmod{4}, 
\\
2(i_2 +k_2)-n_2 &= 0 \pmod{2}. 
\end{align}
\esubeq
Then the sectors with $n_1=0,2$ and $n_2=0$ contribute to the amplitude.
The intersection number is obtained in the same way as in the case of annulus.
In Table \ref{z4z2-fix-a}, we can see that 
$\chi_{\cM}=1$ for untwisted sectors
and $\chi_{\cM}=2$ for $\theta^2$-twisted sectors.

The momentum modes are evaluated in a similar way of subsection \ref{42Klein},
however the winding modes are changed due to the insertions.
In the untwisted sector with the $\OR\theta$ insertion,
from the condition (\ref{m-condition})
the open string states 
$[(1,0)(3,0)]$, $[(1,1)(3,1)]$, $[(1,1)(3,1)]$, $[(1,1)(3,1)]$,
$[(3,0)(1,0)]$, $[(1,0)(3,0)]$ and $[(1,0)(3,0)]$
contribute to the amplitude.
For instance, in the open string state $[(1,0)(3,0)]$,
the momentum modes, 
which are generated by the dual lattice 
$\Lambda_{\cR\theta,\inv} \cap \Lambda_{\cR\theta^3,\inv}$ with
  its basis $\{ {\bf 2e_5} \}$, are given as
\beq
{\bf p} = \frac{n}{\sqrt{2}} {\bf e}_5.
\eeq
The winding modes invariant under $-\OR\theta$ are given by
\beq
{\bf w} = \frac{2n}{\sqrt{2}} {\bf e}_6.
\eeq
This can also read from
$\Lambda_{-\cR\theta,\inv} \cap \Lambda_{-\cR\theta^3,\inv}$.
The zero mode contribution for this state is represented as $\cL_{1,4}$.

We should take it account of the orientifold actions to the Chan-Paton factors. 
For the open strings $[(i_1,i_2)(i_1-n_1,i_2-n_2)]$, 
the $\OR\theta^{k_1}\phi^{k_2}$-insertion contributes 
in the amplitude as 
\beq
\tr \Big[ (\gamma^{(i_1,i_2)}_{\OR k_1k_2})^{-1} 
(\gamma^{(i_1-n_1,i_2-n_2)}_{\OR k_1k_2})^T \Big]. 
\eeq  
Since only the sectors with $n_1=0,2$ and $n_2=0$ contribute to the amplitude, 
we abbreviate 
\bea
a_{k_1k_2}^{(n_1)} &\equiv& 
\tr \Big[ (\gamma^{(2i_1+1,i_2)}_{\OR k_1k_2})^{-1} 
(\gamma^{(2i_1+1+n_1,i_2)}_{\OR k_1k_2})^T \Big], \\
b_{k_1k_2}^{(n_1)} &\equiv& 
\tr \Big[ (\gamma^{(2i_1,i_2)}_{\OR k_1k_2})^{-1} 
(\gamma^{(2i_1+n_1,i_2)}_{\OR k_1k_2})^T \Big]. 
\eea 
These assignments correspond to 
two different classes of the D-brane configurations in this model, 
and are sufficient to evaluate the tadpole cancellation conditons 
for $\Z_4 \times \Z_2$ models. 
However we will need more independent variables 
for $\Z_2 \times \Z_2$ models.

For the contributions from untwisted sector, 
we can use the results from the (\ref{k-mw}) and (\ref{a-mw}) 
owing to the relations (\ref{MK})-(\ref{WA}).  

To summarize, we obtain the M\"obius strip amplitude in the loop channel as
\begin{align}
    {\cal M} = - \frac{N c}{4}(1_{\text{RR}}-1_{\text{NSNS}})
    &\int_{0}^{\infty}
    \frac{dt}{t^3} 
     \Big( \frac{a_{0,0}^{(0)}+b_{0,0}^{(0)}}{2} \cL_\cM^{(0,0)} \cM^{(0,0)(0,0)}
    + 2\frac{a_{3,0}^{(2)}+b_{3,0}^{(2)}}{2}\cL_{1,4}\cM^{(2,3)(0,0)}\nn\\
    &\LS + \frac{a_{2,0}^{(0)}+b_{2,0}^{(0)}}{2}\cL_{1,4}\cM^{(0,2)(0,0)} 
    + 2 \frac{a_{1,0}^{(2)}+b_{1,0}^{(2)}}{2}\cL_{1,4}\cM^{(2,1)(0,0)}\nn\\
    &\LS + \frac{a_{0,1}^{(0)}\cL_{2,2}+b_{0,1}^{(0)}\cL_{1,4}}{2}\cM^{(0,0)(0,1)}
    + 2 \frac{a_{3,1}^{(2)}+b_{3,1}^{(2)}}{2}\cM^{(2,3)(0,1)} \nn\\
    &\LS + \frac{a_{2,1}^{(0)}\cL_{1,4}+b_{2,1}^{(0)}\cL_{2,2}}{2}\cM^{(0,2)(0,1)}
    + 2 \frac{a_{1,1}^{(2)}+b_{1,1}^{(2)}}{2}\cM^{(2,1)(0,1)} \Big).
\end{align}
The modular transformation to the tree channel yields
\begin{align}
    \ctM = -4 c(1_{\text{RR}}-1_{\text{NSNS}})
    &\int_{0}^{\infty} dl 
     \Big( \frac{a_{0,0}^{(0)}+b_{0,0}^{(0)}}{2}\ctL_\cM^{(0,0)} \ctM^{(0,0)}
    - (a_{3,0}^{(2)}+b_{3,0}^{(2)})\ctL_{8,2}\ctM^{(1,0)} \nn\\
    &\LS + 2(a_{2,0}^{(0)}+b_{2,0}^{(0)})\ctL_{8,2}\ctM^{(2,0)} 
    - (a_{1,0}^{(2)}+b_{1,0}^{(2)})\ctL_{8,2}\ctM^{(3,0)} \nn\\
    &\LS + 2(a_{0,1}^{(0)}\ctL_{4,4}+b_{0,1}^{(0)}\ctL_{8,2})\cM^{(0,1)}
    + 2(a_{3,1}^{(2)}+b_{3,1}^{(2)})\ctM^{(1,1)} \nn\\
    &\LS + 2(a_{2,1}^{(0)}\ctL_{8,2}+b_{2,1}^{(0)}\ctL_{4,4})\ctM^{(2,1)}
    + 2(a_{1,1}^{(2)}+b_{1,1}^{(2)})\ctM^{(3,1)} \Big).
\label{Z4Z2-ctM}
\end{align}
To obtain the complete projector and to cancel the tadpole \cite{Forste:2000hx}, 
we set
\bea
&a_{0,0}^{(0)} =a_{1,0}^{(2)} =-a_{2,0}^{(0)} =a_{3,0}^{(2)} =
-a_{0,1}^{(0)} =-a_{1,1}^{(2)} =-a_{2,1}^{(0)} =a_{3,1}^{(2)} =N, \\
&b_{0,0}^{(0)} =b_{1,0}^{(2)} =-b_{2,0}^{(0)} =b_{3,0}^{(2)} =
-b_{0,1}^{(0)} =-b_{1,1}^{(2)} =-b_{2,1}^{(0)} =b_{3,1}^{(2)} =N. 
\label{cp-condition}
\eea

Let us focus on the coefficients on the zero mode contributions in the
Klein bottle amplitude (\ref{Z4Z2-ctK}), the annulus amplitude
(\ref{Z4Z2-ctA}) and the M\"{o}bius strip amplitude (\ref{Z4Z2-ctM}). 
The RR-tadpole cancellation condition (\ref{RR-cancel}) leads to 
\begin{align}
0 &= 16 + \frac{N^2}{4} - 4 N 
= \frac{1}{4} (N - 8)^2
.
\end{align}
The number of one stack of the D-branes is $N=8$ 
to cancel the RR-tadpole. 
Taking account of (\ref{z2-condition}) and (\ref{cp-condition}), 
the gauge groups are determined as $(Sp(2))^4$ for the {\bf DCC} model. 

For the {\bf CCC} model, one of whose orientifold actions is
given by
\beq
\cR = (\ba,\ba,\ba).
\eeq
On the other hand, the element $\cR\theta$ in the orientifold group is
given by
\beq
\cR\theta = (-\bb,\bb,\ba).
\eeq
As seen in Table \ref{o-action},
these two elements yield different numbers of O-planes.
To show this, we evaluate the RR-tadpole amplitude in the following way:
In the tree channel the Klein bottle amplitude is
\begin{align}
    \ctK = c(1_{\text{RR}}-1_{\text{NSNS}})
    &\int_{0}^{\infty} dl 
     \Big( 20\ctL_\cK^{(0,0)} \ctK^{(0,0)}
    - 32\ctL_{2,8}\ctK^{(1,0)} \nn\\
    &\LS - 80\ctL_{2,8}\ctK^{(2,0)} 
    - 32\ctL_{2,8}\ctK^{(3,0)} \nn\\
    &\LS - 40(\ctL_{2,8}+\ctL_{1,4})\cK^{(0,1)}
    + 64\ctK^{(1,1)} \nn\\
    &\LS - 40(\ctL_{2,8}+\ctL_{1,4})\ctK^{(2,1)}
    - 64\ctK^{(3,1)} \Big) .
\label{Z4Z2-ctK-CCC}
\end{align}
The prefactors do not correspond to that from the complete projector.
The annulus and the M\"{o}bius strip amplitudes are also described as
\bsubeq
\begin{align}
    \ctA = \frac{c}{16} (1_{\text{RR}}-1_{\text{NSNS}})
    &\int_{0}^{\infty} dl 
     \Big( (M^2+4N^2) \ctL_\cA^{(0,0)} \ctA^{(0,0)}
    - 8MN\ctL_{2,2}\ctA^{(1,0)} \nn\\
    &\LS - 4(M^2+4N^2)\ctL_{2,2}\ctA^{(2,0)}
    - 8MN\ctL_{2,2}\ctA^{(3,0)} \nn\\
    &\LS - 4(M^2\ctL_{2,2}+4N^2\ctL_{1,4})\cA^{(0,1)}
    + 16MN\ctA^{(1,1)} \nn\\
    &\LS - 4(M^2\ctL_{2,2}+4N^2\ctL_{1,4})\ctA^{(2,1)}
    - 16MN\ctA^{(3,1)} \Big), \label{Z4Z2-ctA-CCC} \\
    \ctM = - c(1_{\text{RR}}-1_{\text{NSNS}})
    &\int_{0}^{\infty} dl 
     \Big( 2(M+N)\ctL_\cM^{(0,0)} \ctM^{(0,0)}
    - 2(M+4N)\ctL_{8,2}\ctM^{(1,0)} \nn\\
    &\LS - 8(M+N)\ctL_{8,2}\ctM^{(2,0)}
    - 2(M+4N)\ctL_{8,2}\ctM^{(3,0)} \nn\\
    &\LS - 8(M\ctL_{8,2}+N\ctL_{4,4})\cM^{(0,1)}
    + 4(M+4N)\ctM^{(1,1)} \nn\\
    &\LS - 8(M\ctL_{8,2}+N\ctL_{4,4})\ctM^{(2,1)}
    - 4(M+4N)\ctM^{(3,1)} \Big) , \label{Z4Z2-ctM-CCC}
\end{align}
\esubeq
where $M$ and $N$ are the numbers of D-branes which are invariant
under the set of orientifold actions $\{ \cR, \cR \theta^2, \cR
\phi, \cR\theta^2 \phi \}$, and under the other set of actions  
$\{ \cR\theta, \cR\theta^3, \cR \theta \phi, \cR \theta^3 \phi\}$,
respectively.  
In the M\"obius strip amplitude we have set 
\bea
&a_{0,0}^{(0)} =a_{1,0}^{(2)} =-a_{2,0}^{(0)} =a_{3,0}^{(2)} =
-a_{0,1}^{(0)} =-a_{1,1}^{(2)} =-a_{2,1}^{(0)} =a_{3,1}^{(2)} =M, \\
&b_{0,0}^{(0)} =b_{1,0}^{(2)} =-b_{2,0}^{(0)} =b_{3,0}^{(2)} =
-b_{0,1}^{(0)} =-b_{1,1}^{(2)} =-b_{2,1}^{(0)} =b_{3,1}^{(2)} =N. 
\eea

Focus on the coefficient in (\ref{Z4Z2-ctK-CCC}),
(\ref{Z4Z2-ctA-CCC}) and (\ref{Z4Z2-ctM-CCC}), we obtain 
the RR-tadpole cancellation conditions (\ref{RR-cancel}), 
\bsubeq
\begin{align}
0 &= 20 + \frac{1}{16} (M^2 + 4 N^2) - 2 (M + N)
= \frac{1}{16} \Big( (M-16)^2 + (N-4)^2 \Big)
, \\
0 &= -32 - \frac{MN}{2} + 2 (M + 4N)
= - \half (M - 16)(N-4) ,
\end{align}
\esubeq
and find $M = 16$ and $N=4$.
This indicates that 
we should insert sets of different numbers of D-branes in an
appropriate way in several kinds of non-factorizable tori.

The open string massless spectrum is given in Table \ref{z4z2-open}.
The multiplicities of twisted states spectra depend on 
the intersection numbers \cite{Forste:2000hx} (see Table \ref{z4z2-fix-a}). 
We see that the {\bf CCD} and {\bf DCC} models are distinct 
from the {\bf CDD} model despite the same numbers of O-planes, 
and actually these four models have different spectra. 
For the closed string the numbers of massless states 
are considerably reduced 
due to their Hodge numbers in \cite{Forste:2006wq,Takahashi:2007qc}. 
\begin{table}[htb]
\renewcommand{\arraystretch}{1.3}
  \begin{center}
      \begin{tabular}{|c||c|c|c|c||c|} \hline
        sectors   & {\bf CCC} & {\bf CCD} 
        & {\bf CDD} &  {\bf DCC} & representations \\
\hline\hline
\multirow{3}{*}{untwisted} & \multicolumn{4}{c||}{$1V$}  
& $Sp[M/4]^2 \times Sp[N/4]^2$ \\ \cline{2-6}
& \multicolumn{4}{c||}{\multirow{2}{*}{$3C$}}  
& $(\Yasym ,1;1,1) \oplus (1,\Yasym;1,1)$ \\ 
& \multicolumn{4}{c||}{ }  
& $\oplus (1,1;\Yasym,1) \oplus (1,1;1,\Yasym)$  \\
\hline
$\theta + \theta^3$   & $2C$ & $2C$ & $4C$ &  $2C$ 
& $(\fund,\fund;1,1) \oplus (1,1;\fund,\fund)$ \\
\hline
\multirow{2}{*}{$\theta^2$}   & $4C$ & $2C$ & $4C$ &  $2C$ 
& $(\Yasym,1;1,1) \oplus (1,\Yasym;1,1)$ \\ \cline{2-6}
                   & $1C$ & $2C$ & $4C$ &  $2C$ 
&  $(1,1;\Yasym,1) \oplus (1,1;1,\Yasym)$ \\
\hline
\multirow{2}{*}{$\phi$} & $4C$ & $2C$ & $4C$ & $2C$ 
& $(\fund,1;\fund,1)$ \\ \cline{2-6}
               & $1C$ & $2C$ & $4C$ & $2C$ 
& $(1,\fund;1,\fund)$ \\
\hline
$\theta\phi + \theta^3\phi$   & $2C$ & $2C$ & $4C$ & $2C$ 
& $(\fund,1;1,\fund) \oplus (1,\fund;\fund,1)$ \\
\hline
\multirow{2}{*}{$\theta^2\phi$}   & $4C$ & $2C$ & $4C$ & $2C$ 
& $(\fund,1;\fund,1)$ \\ \cline{2-6}
                       & $1C$ & $2C$ & $4C$ & $2C$ 
& $(1,\fund;1,\fund)$ \\
\hline
      \end{tabular}
  \end{center}
  \caption{\it Open string massless spectra of 
${\mathbb Z}_4 \times {\mathbb Z}_2$ orbifold on the $D_6$ lattice.
The symbols ``$V$'' and ``$C$'' denote the vector and chiral
  multiplets, respectively.}
 \label{z4z2-open}
\end{table}


\subsection{${\mathbb Z}_2 \times {\mathbb Z}_2$ model}

Since ${\mathbb Z}_2 \times {\mathbb Z}_2$ is a subgroup of 
${\mathbb Z}_4 \times {\mathbb Z}_2$, 
the calculation is similar to the examples in the previous subsection. 
The new feature in ${\mathbb Z}_2 \times {\mathbb Z}_2$ 
is that we have more freedom to choose orbifold actions 
in comparison with the case of ${\mathbb Z}_4 \times {\mathbb Z}_2$. 

For ${\mathbb Z}_2 \times {\mathbb Z}_2$ orbifolds 
on the $D_6$ lattice (\ref{d6-lattice}), 
all the point group elements can be given 
by the use of $\ba$ and $\bb$ in (\ref{ab10}), 
see Appendix \ref{App-LRL}. 
In the case of the {\bf CCC} orientifold with $\cR =(\ba,\ba,\ba)$,  
the point group elements $\theta$ and $\phi$ are 
\bea
\theta:~ (-\b1 ,-\b1 ,\b1 ), \ls
\phi:~ (\b1 ,-\b1 ,-\b1 ). 
\eea
The orientifold group elements including $\OR$ are
\beq
\{ \OR,~ \OR\theta,~ \OR\phi,~ \OR\theta\phi \},
\eeq
and these elements generate O6-planes respectively. 
From Table \ref{z4z2d6} the numbers of O6-planes are read two for each elements. 

In the {\bf CCD} orientifold with $\cR =(\ba,\ba,\bb)$,  
we have two distinct pairs of
the point group elements: 
\begin{align}
\left\{
\begin{array}{r@{: \ }l}
\theta & (\b1 , -\b1 , -\b1 ) \\
\phi & (-\b1 , - \b1 , \b1 )
\end{array} \right.
\ls
\left\{
\begin{array}{r@{: \ }l}
\theta & ( \b1 , -\b1 , -\b1 ) \\
\phi & ( -\b1 , - \ba , \bb )
\end{array} \right.
\end{align} 
The numbers of O-planes generated by the former orbifold actions are also two. 
In the latter case, 
the $\OR$ and $\OR\theta$ ($\OR\phi$ and $\OR\theta\phi$) 
generate two (four) O6-planes, respectively. 
We can classify the distinct orientifold models on the Lie root lattices, 
and the other possible elements on the $D_6$ lattice 
are listed in Table \ref{z2z2d6}. 
We should notice that 
even though the numbers of O6-planes are the same in any three-cycles 
in $\Z_2 \times \Z_2$ orientifold models, 
those of non-factorizable models can be different. 
\begin{table}[htb]
\renewcommand{\arraystretch}{1.3}
\begin{center}
 \begin{tabular}[t]{|c|c|c|c|c|p{9.5mm}|p{9.5mm}|p{9.5mm}|p{9.5mm}|} \hline
  \multirow{2}{*}{Lattice}
  & \multirow{2}{*}{Label}
  & \multirow{2}{*}{reps.\,of $\cR$}
  & \multicolumn{2}{|c|}{Orbifold}
  & \multicolumn{4}{|c|}{\# of O6-planes} \\
  \cline{4-9}
  & & & rep. of $\theta$ & rep. of $\phi$ 
& \hfil $\cR$ & \hfil $\cR\theta$ & \hfil $\cR\phi$ & \hfil $\cR\theta\phi$ \\
  \hline\hline
  \multirow{10}{*}{$D_6$}
  & \multirow{1}{*}{${\bf CCC}$}
      & \multirow{1}{*}{$(\bf{a},\bf{a},\bf{a})$}
	  & $(\bf{1},-\bf{1},-\bf{1})$
	      & $(-\bf{1},-\bf{1},\bf{1})$
		  & \hfil $4$ & \hfil $4$ & \hfil $4$ & \hfil $4$ \\
  \cline{2-9}
  & \multirow{2}{*}{${\bf CCD}$}
      & \multirow{2}{*}{$(\bf{a},\bf{a},\bf{b})$}
	  & $(\bf{1},-\bf{1},-\bf{1})$
	      & $(-\bf{1},-\bf{1},\bf{1})$
		  & \hfil $2$ & \hfil $2$ & \hfil $2$ & \hfil $2$ \\
  \cline{4-9}
  &&
	  & $(\bf{1},-\bf{1},-\bf{1})$
	      & $(-\bf{1},-\bf{a},\bf{b})$
		  & \hfil $2$ & \hfil $2$ & \hfil $4$ & \hfil $4$ \\
  \cline{2-9}
  & \multirow{4}{*}{${\bf CDD}$}
      & \multirow{4}{*}{$(\bf{a},\bf{b},\bf{b})$}
	  & $(\bf{1},-\bf{1},-\bf{1})$
	      & $(-\bf{1},-\bf{1},\bf{1})$
		  & \hfil $1$ & \hfil $1$ & \hfil $1$ & \hfil $1$ \\
  \cline{4-9}
  &&
	  & $(\bf{1},-\bf{1},-\bf{1})$
	      & $(-\bf{1},\bf{b},-\bf{b})$
		  & \hfil $1$ & \hfil $1$ & \hfil $4$ & \hfil $4$ \\
  \cline{4-9}
  &&
	  & $(-\bf{1},\bf{1},-\bf{1})$
	      & $(\bf{a},-\bf{1},-\bf{b})$
		  & \hfil $1$ & \hfil $1$ & \hfil $2$ & \hfil $2$ \\
  \cline{4-9}
  &&
	  & $(\bf{a},-\bf{1},-\bf{b})$
	      & $(-\bf{a},\bf{b},-\bf{1})$
		  & \hfil $1$ & \hfil $2$ & \hfil $2$ & \hfil $4$ \\
  \cline{2-9}
  & \multirow{3}{*}{${\bf DDD}$}
      & \multirow{3}{*}{$(\bf{b},\bf{b},\bf{b})$}
	  & $(-\bf{1},-\bf{1},\bf{1})$
	      & $(\bf{1},-\bf{1},-\bf{1})$
		  & \hfil $2$ & \hfil $2$ & \hfil $2$ & \hfil $2$ \\
  \cline{4-9}
  &&
	  & $(\bf{1},-\bf{1},-\bf{1})$
	      & $(-\bf{1},-\bf{b},\bf{b})$
		  & \hfil $2$ & \hfil $2$ & \hfil $2$ & \hfil $2$ \\
  \cline{4-9}
  &&
	  & $(-\bf{1},-\bf{b},\bf{b})$
	      & $(\bf{b},-\bf{1},-\bf{b})$
		  & \hfil $2$ & \hfil $2$ & \hfil $2$ & \hfil $2$ \\
  \hline
 \end{tabular}
\caption{\it $\Z_2 \times \Z_2$ orbifold models on the $D_6$ Lie root lattice.}
\label{z2z2d6}
\end{center}
\end{table}
%

Finally we check the RR-tadpole cancellation 
in the $\Z_2 \times \Z_2$ {\bf CCC} model on the $D_6$ lattice.
The contribution from $\phi$- and $\theta \phi$-twisted sectors are the same as
$\theta$-sector for the {\bf CCC} model on the $D_6$ lattice. 
The RR-tadpole cancellation is satisfied with $N=4$ 
as we can see the following amplitudes in the tree channel. 
The Klein bottle amplitude is given as
\begin{align}
    \ctK = 32c(1_{\text{RR}} - 1_{\text{NSNS}}) 
    \int_{0}^{\infty} dl 
     \Big( \ctL_\cK^{(0,0)} \ctK^{(0,0)} 
    - 4\ctL_{2,8}\ctK^{(1,0)} 
    - 4\ctL_{2,8}\ctK^{(0,1)} 
    - 4\ctL_{2,8}\ctK^{(1,1)} \Big).
\end{align}
The annulus and the M\"{o}bius amplitudes are also given as
\bsubeq
\begin{align}
    \ctA &= \frac{N^2 c}{8} (1_{\text{RR}} - 1_{\text{NSNS}}) 
    \int_{0}^{\infty} dl 
     \Big( \ctL_\cA^{(0,0)} \ctA^{(0,0)} 
    - 4\ctL_{2,2}\ctA^{(1,0)} 
    - 4\ctL_{2,2}\ctA^{(0,1)} 
    - 4\ctL_{2,2}\ctA^{(1,1)} \Big) ,
\\
    \ctM &= -4N c(1_{\text{RR}}-1_{\text{NSNS}}) 
    \int_{0}^{\infty} dl 
     \Big( \ctL_\cM^{(0,0)} \ctM^{(0,0)} 
    - 4\ctL_{8,2}\ctM^{(1,0)} 
    - 4\ctL_{8,2}\ctM^{(0,1)} 
    - 4\ctL_{8,2}\ctM^{(1,1)} \Big).
\end{align}
\esubeq
We observe that in any amplitudes 
the prefactors are given by the complete projector (\ref{complete}).

\section{Conclusion}
\label{Conclusion}

In this paper we studied the RR-tadpole cancellation condition
in Type II string models compactified on six-tori given by general Lie
root lattices. 
We obtained a simple derivation to count the orientifold planes lying on
the lattice 
by the use of the Lefschetz fixed point theorem. 
As expected the RR-tadpole contributions are cancelled 
by adding an appropriate number of D-branes parallel to the O-planes. 
The Lefschetz fixed point theorem provides
an intuitive picture to non-factorizable models, 
and we easily showed a way to 
construct orientifold models on tori and orbifolds. 

In $D=4$, ${\cal N}=1$
$\Z_N \times \Z_M$ orientifolds,
mainly the factorizable models on $T^2 \times T^2 \times T^2$ 
have been constructed and investigated.
We gave the classifications in
Type IIA orientifold models with O6-planes, and obtained many new models. 
As explained in detail, 
the Lefschetz fixed point theorem provide 
intuitive and convenient tools in model construction. 
Since the condition derived in (\ref{o6-condition}) is 
the necessary condition for orbifolds, 
we performed explicit calculations 
for $\Z_4 \times \Z_2$ and $\Z_2 \times \Z_2$ orbifold models, 
and confirmed the RR-tadpole calculations. 
It is expected that even in other non-factorizable orbifold models 
the RR-tadpole cancellation should be checked in the same calculation.
We further found many non-factorizable $\Z_2 \times \Z_2$
orbifolds in which the numbers of O-planes depend on the three-cycles
left invariant under the orbifold projections in Table \ref{z2z2d6} 
and in Table \ref{z2z2a3a3}. 
These features are not seen in factorizable models, 
and will provide new possibilities for model constructions. 
On the other hand,
since the metric of non-factorizable tori is 
changed to $B$-field via T-duality, 
our consideration should be related to compactification with such backgrounds. 
Actually in heterotic orbifolds 
there are some coincidences between non-factorizable models and 
factorizable models with generalized discrete torsion 
\cite{Ploger:2007iq}. 
Our results indicate that there would be a possibility to construct various 
class of $D=4$, ${\cal N}=1$ models 
with different set of chiral spectra from 
other well-known (non-)factorizable models.

\subsection*{Acknowledgements}

K.T. is supported by the Grand-in-Aid for Scientific Research \#172131. 
T.K. is supported by the Grant-in-Aid for the 
21st Century COE ``{\sc Center for Diversity and Universality in Physics}'' 
from the Ministry of 
Education, Culture, Sports, Science and Technology (MEXT) of Japan.

\appendix

\section*{Appendix}

\section{Six-dimensional Lie root lattice}
\label{App-LRL}

In this appendix we study basic aspects of the Lie root lattice 
given by the simple Lie algebra 
and its application to non-factorizable six-tori. 
First we review the simple root on the Lie algebra. 
In terms of Lie root lattices, 
we find that there are only twelve distinct non-factorizable six-tori 
and four factorizable ones. 
By the use of the Weyl reflection and the outer automorphisms,
we can classify all the point groups of orbifolds and 
orientifold actions $\cR$ on the tori, 
which crystallographically act on the Lie root lattices. 
We give explicit representations of 
point group elements generated by the Weyl reflections and 
the outer automorphisms of the Lie root lattices. 
Some of the point groups can be given by 
the Coxeter elements from the Cater diagrams 
or the generalized Coxeter elements as explained later.
Beside these elements, 
we see that point groups which are not included 
in the (generalized) Coxeter elements 
are also obtained by the classification.  

We utilize these elements for the point groups of our orientifold models, 
and these elements lead to many new orientifold models 
as explained in section \ref{ZZorient} and in this appendix.
Here we give the systematic way to construct 
orbifolds and orientifolds on the Lie root lattices.


\subsection{Lie root lattices}

We use the words of the Lie algebra 
in order to define the shape of tori defined in (\ref{def-torus}). 
The Lie algebras whose orders are within six 
are $A_N$, $B_N$, $C_N$, $D_N$, $E_6$, $F_4$ and $G_2$. 
The simple roots ${\bf \alpha}_i$ of these Lie algebras
can be given as follows: 
\begin{align}
\begin{array}{l@{\; \ \ }ll}
A_N: & {\bf \alpha}_i = {\bf e}_{i}-{\bf e}_{i+1} , & i=1, \dots, N  \\
B_N: & {\bf \alpha}_i = {\bf e}_{i}-{\bf e}_{i+1} , \; \; {\bf \alpha}_N = {\bf e}_{N} 
, & i=1, \dots, N-1  \\
C_N: & {\bf \alpha}_i = {\bf e}_{i}-{\bf e}_{i+1} , \; \;  {\bf \alpha}_N = 2{\bf e}_{N}
, & i=1, \dots, N-1  \\
D_N: & {\bf \alpha}_i = {\bf e}_{i}-{\bf e}_{i+1} ,  \; \; {\bf \alpha}_N = {\bf e}_{N-1} +{\bf e}_{N}  
, & i=1, \dots, N-1 \\
E_6: & {\bf \alpha}_i = {\bf e}_{i}-{\bf e}_{i+1} ,  \; \; 
{\bf \alpha}_5 = {\bf e}_{4}+{\bf e}_{5} ,  & i=1, \dots, 4 \\
 & {\bf \alpha}_6 = -\frac12 ({\bf e}_{1}+{\bf e}_{2}+{\bf e}_{3}+{\bf e}_{4}+{\bf e}_{5}+ \sqrt{3} {\bf e}_{6})   &
\\
F_4: & {\bf \alpha}_1 = {\bf e}_{1}-{\bf e}_{2} ,  \; \;  \; \;{\bf \alpha}_2 ={\bf e}_{2}-{\bf e}_{3} ,   \; \;
{\bf \alpha}_3 =2{\bf e}_{3} , &
{\bf \alpha}_4 =-{\bf e}_{1}-{\bf e}_{2}-{\bf e}_{3}-{\bf e}_{4} 
\\
G_2: & {\bf \alpha}_1 = {\bf e}_{1}-{\bf e}_{2} ,  \; \;  \; \;{\bf \alpha}_2 = -{\bf e}_{1} +2{\bf e}_{2} -{\bf e}_{3} , & 
\label{simple-roots}
\end{array}
\end{align}
where ${\bf e}_i$'s are unit vectors 
whose scalar product is defined as 
${\bf e}_i \cdot {\bf e}_j = \delta_{ij}$.
The Dynkin diagrams are drawn in Figure \ref{dynkins}.
\begin{figure}[tb]
\begin{center}
\includegraphics{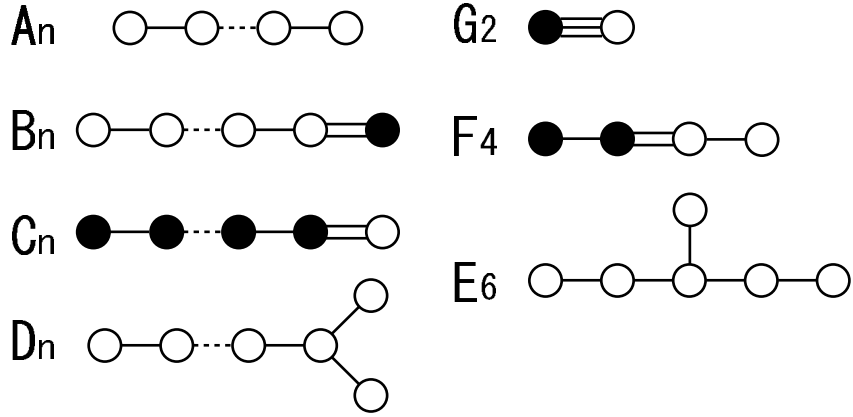}
\caption{\it Dynkin diagrams of the simple Lie algebras.}
\label{dynkins}
\end{center}
\end{figure}
From these diagrams 
one can easily find a set of equivalence relations (isomorphism) among
the simple Lie algebra, 
\begin{align}
A_1 \sim B_1 \sim C_1, \ls 
B_2 \sim C_2 , \ls A_3 \sim D_3 .
\end{align}
We further find the equivalence relations {\it from the Lie root lattice
point of view}: 
\bsubeq
\begin{gather}
A_2 \sim G_2 , \ls
B_2 \sim D_2 \sim (A_1)^2 
, \ls
D_4 \sim F_4 
, \\
B_N \sim (A_1)^N , \ls
C_N \sim D_N  ,
\end{gather}
\esubeq
where $(A_1)^2 = A_1 \times A_1$.
Here we assumed the most symmetric cases, 
where the lengths of the shortest roots are equal 
between the lattices given as direct products. 
Since we are interested in symmetries of the lattices, 
the assumption is rational. 
We often use these equivalence relations 
in the classification of the six-tori. 

Taking the direct products of tori generated from these lattices, 
we obtain six-tori in terms of the Lie root lattices.
We conclude that there are only twelve inequivalent non-factorizable six-tori 
and four factorizable ones\footnote{Most of other six-tori 
would be obtained by the continuous deformation 
of moduli of these tori \cite{Forste:2006wq}.} 
in such a way as in Table \ref{all-lattices}.
\begin{table}[htb]
\renewcommand{\arraystretch}{1.2}
\normalsize
\begin{center}
\begin{tabular}{l@{\hspace{7mm}}l@{\hspace{7mm}}l@{\hspace{7mm}}l}
\hline
\multicolumn{3}{l}{[non-factorizable tori]}  \\
$A_6$ & $D_6$ & $E_6$ \\
$A_5 \times A_1$ & $A_4 \times A_2$ & $A_4 \times (A_1)^2$ \\
$D_5 \times A_1$ & $D_4 \times A_2$ & $D_4 \times (A_1)^2$ \\
$A_3 \times A_3$ & $A_3 \times A_2 \times A_1$ & $A_3 \times (A_1)^3$ \\
\hline
\multicolumn{3}{l}{[factorizable tori]} \\
$(A_2)^3$ & $(A_2)^2 \times (A_1)^2$ & $A_2 \times (A_1)^4$ & $(A_1)^6$ \\
\hline
\end{tabular}
\end{center}
\caption{\it All the Lie root lattices in six dimensions.}
\label{all-lattices}
\end{table}

\subsection{Weyl reflection and graph automorphism}

Next we investigate the automorphisms of the above lattices. 
Both orbifold and orientifold groups should 
crystallographically act on the lattices. 
These groups can be classified in terms of the Weyl reflection and
the graph automorphism acting on the simple roots of the Lie root lattice.
The Weyl group $\cW$ is generated by the following Weyl reflections 
$r_{{\bf \alpha}_{k}}$ 
which associate the simple root ${\bf \alpha}_{k}$:
\begin{align}
r_{{\bf \alpha}_k}: \ 
\lambda 
\ &\raw \ 
\lambda -2 \frac{{\bf \alpha}_k \cdot \lambda}{|{\bf \alpha}_k|^2} {\bf \alpha}_k
.
\label{weyl-ref}
\end{align}
In the case of the $D_N$ Lie root lattice, for instance,
the Weyl reflection $r_{{\bf \alpha}_{k}}$ for $k=1,\dots,N-1$
is given as
\bsubeq
\begin{align}
r_{{\bf \alpha}_{k}} : \ \left\{
\begin{array}{rcl}
{\bf \alpha}_{k-1} 
\ &\raw & {\bf \alpha}_{k-1} + {\bf \alpha}_{k}
\\
{\bf \alpha}_{k} 
\ &\raw & - {\bf \alpha}_{k} 
\\
{\bf \alpha}_{k+1} 
\ &\raw & 
{\bf \alpha}_{k+1} + {\bf \alpha}_{k}
\end{array} \right.
\end{align}
and ${\bf \alpha}_{m}$ for $m \neq k-1,k,k+1$ are unchanged.
For $k=N$ it is 
\begin{align}
r_{{\bf \alpha}_{N}} : \ \left\{
\begin{array}{rcl}
{\bf \alpha}_{N-2} &\raw& {\bf \alpha}_{N-2} + {\bf \alpha}_{N} \\
{\bf \alpha}_{N} &\raw& -{\bf \alpha}_{N} 
\end{array} \right.
\end{align}
\esubeq
and  the other ${\bf \alpha}_{m}$'s are unchanged.
For the classification of the automorphisms, 
it would be convenient to rewrite them 
in the basis of orthogonal unit vectors ${\bf e}_i$ as 
\bsubeq
\begin{align}
r_{{\bf \alpha}_{k}} : \ \ 
{\bf e}_{k} \ &\leftrightarrow \ {\bf e}_{k+1}  \ls
k = 1,\dots, N-1 
, \\
r_{{\bf \alpha}_{N}} : \ \ 
{\bf e}_{N} \ &\leftrightarrow \ -{\bf e}_{N-1} .
\end{align}
\esubeq

On the other hand,
the outer automorphism $g$ of the Cartan diagram is represented as 
\beq
g: ~{\bf \alpha}_{N-1} \leftrightarrow {\bf \alpha}_{N},
\eeq
and the other simple roots are left unchanged. 
In the unit vector basis, it is 
\beq
g: ~{\bf e}_{N} \raw -{\bf e}_{N}.
\eeq
In terms of ${\bf e}_i$, 
we can easily construct any elements generated from 
$r_{{\bf \alpha}_{k}}$ and $g$.   
For example a product of 
two Weyl reflections which do not commute with each other 
makes up $\bZ_3$ element as  
\beq
r_{{\bf \alpha}_{k}}r_{{\bf \alpha}_{k+1}}:~
{\bf e}_{k} \raw {\bf e}_{k+1} \raw {\bf e}_{k+2} \raw {\bf e}_{k}, ~~~~~(k<N-1).
\eeq
This is the permutation group $S_3$. 
Similarly the Weyl reflections $r_{{\bf \alpha}_{k}}$ for $k=1,\dots, N-1$ generate 
a permutation group $S_{N}$. 
Adding the other elements $r_{{\bf \alpha}_{N}}$ and $g$ to $S_N$, 
the representation of the group is given by permutations with signs 
\beq
{\bf e}_i \raw \pm {\bf e}_j \raw \pm {\bf e}_k \raw \cdots \raw \pm {\bf e}_i.
\eeq
Then the order of the Weyl group $\cW$ and $\{ \cW, g \}$ are 
summarized in Table \ref{weyl-order}:
\begin{table}[htb]
\renewcommand{\arraystretch}{1.3}
\normalsize
\begin{center}
\begin{tabular}{|c||c|c|}
\hline
  & $ \cW $ & $\{ \cW,g \}$ \\
\hline
\hline
$D_N$ & $2^{N-1}N!$ & $2^{N}N!$  	\\
\hline
$A_{N-1}$ & $N!$ &  $2N!$ \\
\hline
\end{tabular}
\end{center}
\caption{\it The order of the Weyl group and the graph automorphism.}
\label{weyl-order}
\end{table}

In the case of the $A_{N-1}$ Lie root lattice, 
the Weyl reflections generate permutation group $S_N$ 
in terms of ${\bf e}_i$. 
Its outer automorphism of the Dynkin diagram 
is given by the following permutation
\begin{align}
g:~{\bf e}_{i} \lraw -{\bf e}_{N+1-i},~~~i=1,\cdots,N
\end{align}
We can always permute $g$ to $g^\prime$ by the elements of $\cW$
such that $g^\prime$ change the sign of all ${\bf e}_{i}$'s, 
This element is expressed as an identity matrix with negative sign $- \id_{N}$, 
which means $\{ \cW,g \}= \{ \cW, - \id_{N} \}$. 
Therefore the order of $\{ \cW,g \}$ is twice as many as that of $\cW$, 
as in Table \ref{weyl-order}.

Then it is straightforward to obtain all $\bZ_2$ elements of the $D_N$ lattice, 
and they are given by the following sub-elements  
\bsubeq
\begin{align}
\begin{split}
{\bf e}_i &\lraw  {\bf e}_j,  \\
{\bf e}_k &\lraw -{\bf e}_l, \\
{\bf e}_m &\raw - {\bf e}_m, 
\end{split}
\end{align}
\esubeq
except for $D_4$. 
The $\bZ_N$ elements are constructed similarly.
For example $\bZ_3$ elements are constructed 
by the following sub-elements,
\beq
{\bf e}_i \raw \pm {\bf e}_j \raw \pm {\bf e}_k \raw \pm {\bf e}_i, 
\ls \text{(\# of terms with $-$ sign is even)}
\eeq
and their permutations. 
$\bZ_4$ elements includes the following sub-elements,
\bsubeq
\begin{gather}
{\bf e}_i \raw - {\bf e}_j \raw - {\bf e}_i, \ls i \neq j, \\
{\bf e}_i \raw \pm {\bf e}_j \raw \pm {\bf e}_k \raw \pm {\bf e}_l 
\raw \pm {\bf e}_i, \ls \text{(\# of terms with $-$ sign is even)}.
\end{gather}
\esubeq
We can similarly deal with the $A_N$ lattices. 
Note that the roots of the $A_N$ and $D_N$ can be given by 
\bsubeq
\bea
A_N: &  {\bf e}_{i}-{\bf e}_{j}  , & \\
D_N: &  \pm{\bf e}_{i}\pm{\bf e}_{j}  , & i,j=1, \dots, N  .
\eea
\esubeq
They are symmetric under the permutations of $i$ and $j$. 
Now it is apparent that on the $D_6$ lattice, 
(\ref{z4z2}) is the only inequivalent $\Z_4 \times \Z_2$ elements,
and $\Z_2 \times \Z_2$ elements can be given 
by $\bf{a}$, $\bf{b}$ and $\bf{1}$ in (\ref{ab10}). 
From Table \ref{all-lattices},
we have all the point group elements which can be expressed 
by the Weyl reflections and the outer automorphism (except for the $E_6$ lattice). 
However there are a few exceptions 
owing to additional outer automorphisms as follows. 

We shortly explain the Coxeter elements and 
the generalized Coxeter elements\footnote{From the definition of the
  (generalized) Coxeter elements, 
we can see that the elements do not left any directions invariant 
for corresponding sub-space. 
Then it is apparent that for $\Z_N \times \Z_M$ orbifold 
they lead to factorizable models on $T^2 \times T^2 \times T^2$.}. 
The Coxeter element of the Lie root lattice 
is defined by product of all the Weyl reflections 
which associate with simple roots, 
\beq
D_N \equiv r_{{\bf \alpha}_1} r_{{\bf \alpha}_2} \cdots r_{{\bf \alpha}_N}.
\eeq
The other Coxeter elements, 
which are generated by different ordering of product, 
are conjugate to one another, 
and lead to the same class of orbifolds.
There are other elements generated by the Weyl reflections.
These orbifolds can be classified 
by the Carter diagrams \cite{Schellekens:1987ij}.
The Coxeter elements of $D_4$ 
from the Carter diagrams are, for example,
\bsubeq
\begin{align}
D_4 &= r_{{\bf \alpha}_1} r_{{\bf \alpha}_2} r_{{\bf \alpha}_3} r_{{\bf \alpha}_4}, \\
D_4(a1) &= r_{{\bf \alpha}_1}r_{{\bf \alpha}_2}r_{{\bf \alpha}_3}
r_{{\bf \alpha}_2+{\bf \alpha}_3+{\bf \alpha}_4},
\end{align}
\esubeq
where $r_{{\bf \alpha}_2+{\bf \alpha}_3+{\bf \alpha}_4}$ is 
a Weyl reflection associated 
with the sum of simple roots ${\bf \alpha}_2+{\bf \alpha}_3+{\bf \alpha}_4$. 
Then the order of $D_4$ is six, and that of $D_4(a1)$ is four.
However these elements do not include 
the outer automorphisms\footnote{There would be complete classifications 
including the outer automorphisms by mathematicians.
However the authors do not know it. 
Alternatively 
our approach provides a complete classification 
and useful formula for 
the six-dimensional Lie root lattices, except for $E_6$.}. 
The generalized Coxeter elements are 
defined by adding outer automorphisms to the Coxeter elements. 
For example the $D_N$ Lie root lattice has 
a graph automorphism $g$ 
which exchanges the simple root ${\bf \alpha}_{N-1}$ and ${\bf \alpha}_{N}$. 
The generalized Coxeter element is defined by  
\beq
C^{[2]} \equiv r_{{\bf \alpha}_1} r_{{\bf \alpha}_2} 
\cdots r_{{\bf \alpha}_{N-2}} g. 
\eeq
For instance the generalized Coxeter element of $D_4$ is
\beq
C^{[2]} = r_{{\bf \alpha}_1} r_{{\bf \alpha}_2} r_{{\bf \alpha}_3} g,
\eeq
and the order of this element is eight.

Actually these (generalized) Coxeter elements and 
elements from the Cater diagrams are 
included in the above classification by the use of ${\bf e}_i$. 
An exception occurs in the $D_4$ lattice,  
which has another outer automorphism $g^\prime$,
\beq 
g^\prime:~{\bf \alpha}_1 \raw {\bf \alpha}_3 
\raw {\bf \alpha}_4 \raw {\bf \alpha}_1. 
\eeq
The generalized Coxeter element of this outer automorphism is defined by 
\beq
C^{[3]} \equiv r_{{\bf \alpha}_1} r_{{\bf \alpha}_2} g^\prime .
\eeq
This action corresponds to 
a rotation of $(e^{\pi i/6},e^{5\pi i/6})$. 
For this element the classification in the ${\bf e}_i$ basis is inconvenient 
(since for example it acts as $g^\prime:~{\bf e}_1 \raw 
({\bf e}_1+{\bf e}_2+{\bf e}_3+{\bf e}_4)/2$).
We comment that among $\Z_N \times \Z_M$ orbifolds 
this element generates new orbifold only for $\Z_3 \times \Z_3$, 
e.g. $(C^{[3]})^4$ is rotation of $(e^{2\pi i/3},e^{2\pi i/3})$ 
and that of $r_{{\bf \alpha}_3}g^\prime$ is $(1,e^{2\pi i/3})$. 
Then a torus on the $D_4 \times A_2$ lattice allows a $\Z_3 \times \Z_3$ orbifold.

In the case that two independent radii of a torus 
on the $A_3 \times A_3$ lattice are equal to each other, 
there is an additional outer automorphism $g_{33}$, 
\beq
g_{33}: {\bf \alpha}_i \lraw \pm {\bf \alpha}_{i+3},
\eeq
where ${\bf \alpha}_i$ is a simple root of the first (second) $A_3$ for $i=1,2,3$ 
($i=4,5,6$). 
From the observation of its eigenvalues, 
these elements do not generate another $\Z_N \times \Z_M$ elements. 
However the orientifold action $\cR$ can be generated from $g_{33}$,
which will be explained in the next subsection. 
Such outer automorphisms also arise in factorizable tori 
including sublattices $(A_2)^n$ and $(A_1)^m$. 
For example, $(A_2)^2$ has an outer automorphism as 
\beq
g_{22}: {\bf \alpha}_i \raw - {\bf \alpha}_i^\prime 
\raw - {\bf \alpha}_i, ~~i=1,2.
\eeq
where ${\bf \alpha}_i$ is a simple root of the first $A_2$ and 
${\bf \alpha}_i^\prime$ is one of the second $A_2$. 
The eigenvalues of this element are $(e^{\pi i/2},e^{\pi i/2})$, 
and generate $\Z_4$ elements. 
In this case the factorizable tori are actually non-factorizable as orbifolds. 

We investigate the other cases similarly, 
and obtain the allowed $\Z_N \times \Z_M$ orbifolds in Table \ref{zmzn-orbifold}.
In the next subsection we will explain the orientifold actions 
which are compatible with these non-factorizable orbifolds. 
\begin{table}[htb]
\renewcommand{\arraystretch}{1.5}
\renewcommand{\baselinestretch}{1.0}
\normalsize
\begin{center}
\begin{tabular}{|c|c|c|c|c|}
\hline
Lie root lattice 
& $\Z_2 \times \Z_2$ 
& $\Z_2 \times \Z_4$ 
& $\Z_3 \times \Z_3$ 
& $\Z_4 \times \Z_4$ 
\\
\hline
\hline
$A_6$ & -- & -- & -- & -- \\
$D_6$ & $\checked$ & $\checked$ & -- & $\checked$ \\
$E_6$ & $\checked$ & $\checked$ & $\checked$ & -- \\
$A_5 \times A_1$ & $\checked$ & -- & -- & -- \\
$D_5 \times A_1$ & $\checked$ & $\checked$ & -- & -- \\
$A_4 \times A_2$ & $\checked$ & -- & -- & -- \\
$A_4 \times (A_1)^2$ & $\checked$ & -- & -- & -- \\
$D_4 \times A_2$ & $\checked$ & $\checked$ & $\checked$ & -- \\
$D_4 \times (A_1)^2$ & $\checked$ & $\checked$ & -- & $\checked$ \\
$A_3 \times A_3$ & $\checked$ & $\checked$ & -- & -- \\
$A_3 \times A_2 \times A_1$ & $\checked$ & -- & -- & -- \\
$A_3 \times (A_1)^3$ & $\checked$ & $\checked$ & -- & -- \\
\hline
$(A_2)^3$ & $\checked$ & $\checked$ & $\checked$ & -- \\
$(A_2)^2 \times (A_1)^2$ & $\checked$ & $\checked$ & -- & -- \\
$A_2 \times (A_1)^4$ & $\checked$ & $\checked$ & -- & -- \\
$(A_1)^6$ & $\checked$ & $\checked$ & -- & $\checked$ \\ 
\hline
\end{tabular}
\end{center}
\caption{\it Classification of six-dimensional (non-)factorizable tori and
  possible $\Z_N \times \Z_M$ orbifold models on them.}
\label{zmzn-orbifold}
\end{table}


\subsection{Orientifolds on non-factorizable orbifolds}

In the previous subsection 
we gave a way to obtain orbifolds on non-factorizable tori. 
In order to preserve supersymmetry, 
the orbifold action 
$\theta:  (z_1,~z_2,~z_3) \raw 
(e^{2\pi i v_1} z_1,~e^{2\pi i v_2} z_2,~e^{2\pi i v_3} z_3)$
should satisfy the equation $v_1 + v_2 + v_3=0$. 
Then only a holomorphic $(3,0)$-form 
$\Omega = dz_1 \wedge dz_2 \wedge dz_3$ 
and a anti-holomorphic $(0,3)$-form 
$\bar{\Omega}$ are left invariant, 
and the other three forms on a six-tori are generally projected out. 
The orientifold action $\cR$ of O6-plane, 
which preserve $\cN=1$ supersymmetry, should act as  
\beq
\cR:  (z_1,~z_2,~z_3) \raw (a\bar{z}_1,~b\bar{z}_2,~c\bar{z}_3),
\label{R-act-zzz}
\eeq
where $a$, $b$ and $c$ are phase factors. 
Then the every orientifold group element including $\cR$ 
generates fixed loci of O6-planes. 

For their classification 
we again use the abbreviations {\bf a}, {\bf b} and {\bf 1} in (\ref{ab10}). 
For the $D_6$ lattice 
we have $\Z_2 \times \Z_2$ elements 
as $\theta= (\bf{-1},\bf{-1},\bf{1})$ and $\phi= (\bf{1},\bf{-1},\bf{-1})$. 
The orientifold actions which are compatible with this orbifold are
\bea
(\pm \bf{a},\pm \bf{a}, \pm \bf{a}),~ 
(\underline{\pm \bf{a},\pm \bf{a}, \pm \bf{b}}),~ 
(\underline{\pm \bf{a},\pm \bf{b}, \pm \bf{b}}),~
(\pm \bf{b},\pm \bf{b}, \pm \bf{b}),
\eea
where the underlined entries are permuted.
For the orbifold elements 
$\theta= (\bf{-1},\bf{a},\bf{b})$, $\phi= (\bf{1},\bf{-1},\bf{-1})$,
the compatible orientifold actions are 
\footnote{Note that for this orbifold elements 
the basis is different from (\ref{R-act-zzz}).}
\bea
\pm(\bf{a}, \bf{a},  -\bf{b}),~ 
\pm(\bf{a}, -\bf{a},  \bf{b}),~ 
\pm(-\bf{b}, \bf{a},  -\bf{b}),~ 
\pm(\bf{b}, -\bf{a},  \bf{b}) .
\eea
In other words, the restriction is that 
the eigenvalues of each orientifold group element 
$\cR$, $\cR\theta$, $\cR\phi$ and $\cR\theta\phi$ should be $(-1,-1,-1,1,1,1)$. 
Note that there are some equivalent actions due to the symmetry of the lattice. 
These considerations lead to Table \ref{z2z2d6} 
for the $\Z_2 \times \Z_2$ orbifold models on the $D_6$ lattice.

There exists an exception in this classification for the $A_3 \times A_3$ lattice 
as mentioned before. 
We define the lattice $A_3 \times A_3$ by using the simple roots
\begin{align}
\begin{array}{c}
{\bf \alpha}_{1} = {\bf e}_{1}-{\bf e}_{2}  , \\
{\bf \alpha}_{2} = {\bf e}_{2} -{\bf e}_{3}  , \\
{\bf \alpha}_{3} = {\bf e}_{2} +{\bf e}_{3}  , 
\end{array} 
\LS
\begin{array}{c}
{\bf \alpha}_{4} = {\bf e}_{4}-{\bf e}_{5}  , \\
{\bf \alpha}_{5} = {\bf e}_{5} -{\bf e}_{6}  , \\
{\bf \alpha}_{6} = {\bf e}_{5} +{\bf e}_{6}  .
\end{array} 
\end{align}
In this base $\Z_2 \times \Z_2$ orbifolds are 
obtained in a similar manner of the $D_6$ lattice\footnote{It may seem
  that the classification with {\bf b},{\bf a} and {\bf 1} elements 
is missing the action 
$\cR:~{\bf \alpha}_{i} \raw -{\bf \alpha}_{i}$ with $i=1,2,3$, however this 
action is included in orientifold groups, e.g. 
the $\cR\theta\phi$ action of {\bf DCD} model on Table \ref{z2z2a3a3}.}. 
Note that the action $\cR=(*, {\bf b}, *)$, where $*$ is 
${\bf b}$, ${\bf a}$ or ${\bf 1}$, 
is forbidden due to the lattice structure. 
The outer automorphism between two $A_3$'s generates
an exceptional action 
\beq
\cR:~{\bf \alpha}_{i} \lraw {\bf \alpha}_{i+3}, ~~i=1,2,3. 
\eeq
If we redefine the base of $A_3 \times A_3$ as 
\begin{align}
\begin{array}{c}
{\bf \alpha}_{1} = {\bf e}_{1}-{\bf e}_{3}  , \\
{\bf \alpha}_{2} = {\bf e}_{3} -{\bf e}_{5}  , \\
{\bf \alpha}_{3} = {\bf e}_{3} +{\bf e}_{5}  , 
\end{array} 
\LS
\begin{array}{c}
{\bf \alpha}_{4} = {\bf e}_{2}-{\bf e}_{4}  , \\
{\bf \alpha}_{5} = {\bf e}_{4} -{\bf e}_{6}  , \\
{\bf \alpha}_{6} = {\bf e}_{4} +{\bf e}_{6}  ,
\end{array} 
\end{align}
the exceptional action is expressed by $(\bf{b}, \bf{b},  \bf{b})$ 
in the orthogonal ${\bf e}_i$ basis: 
\beq
\cR:~{\bf e}_{1} \lraw {\bf e}_{2},~~{\bf e}_{3} \lraw {\bf e}_{4},~~
{\bf e}_{5} \lraw {\bf e}_{6}.
\eeq
Actually this element gives only one inequivalent 
element including the outer automorphism, 
and we label it as $({\bf DDD})^\prime$. 

Including this orientifold action 
we obtain all
the elements of ${\mathbb Z}_2 \times {\mathbb Z}_2$ orbifolds 
on the $A_3 \times A_3$ lattice in Table \ref{z2z2a3a3}.
\begin{table}[htb]
\renewcommand{\arraystretch}{1.2}
\begin{center}
 \begin{tabular}[t]{|c|c|c|c|c|p{9.5mm}|p{9.5mm}|p{9.5mm}|p{9.5mm}|} \hline
  \multirow{2}{*}{Lattice}
  & \multirow{2}{*}{Label}
  & \multirow{2}{*}{rep.\,of $\cR$}
  & \multicolumn{2}{|c|}{Orbifold}
  & \multicolumn{4}{|c|}{\# of O6-planes} \\
  \cline{4-9}
  &&& rep.\,of $\theta$ & rep.\,of $\phi$ 
& \hfil $\cR$ & \hfil $\cR\theta$ & \hfil $\cR\phi$ & \hfil $\cR\theta\phi$ \\
  \hline\hline 
  \multirow{10}{*}{$A_3 \times A_3$}
  & \multirow{2}{*}{${\bf CCC}$}
      & \multirow{2}{*}{$(\bf{a},\bf{a},\bf{a})$}
	  & $(\bf{1},-\bf{1},-\bf{1})$
	      & $(-\bf{1},-\bf{1},\bf{1})$
		  & \hfil $2$ & \hfil $2$ & \hfil $2$ & \hfil $2$ \\
  \cline{4-9}
      && 
	  & $(-\bf{1},\bf{1},-\bf{1})$
	      & $(\bf{a},-\bf{1}, -\bf{a})$
		  & \hfil $2$ & \hfil $2$ & \hfil $2$ & \hfil $8$ \\
  \cline{2-9}
  & \multirow{3}{*}{${\bf CCD}$}
      & \multirow{3}{*}{$(\bf{a},\bf{a},\bf{b})$}
	  & $(\bf{1},-\bf{1},-\bf{1})$
	      & $(-\bf{1},-\bf{1},\bf{1})$
		  & \hfil $2$ & \hfil $2$ & \hfil $2$ & \hfil $2$ \\
  \cline{4-9}
  &&
	  & $(\bf{1},-\bf{1},-\bf{1})$
	      & $(-\bf{1},\bf{a},-\bf{b})$
		  & \hfil $2$ & \hfil $2$ & \hfil $2$ & \hfil $2$ \\
  \cline{4-9}
  &&
	  & $(-\bf{1},\bf{1},-\bf{1})$
	      & $(\bf{a},-\bf{1},-\bf{b})$
		  & \hfil $2$ & \hfil $2$ & \hfil $2$ & \hfil $8$ \\
  \cline{2-9}
  & \multirow{5}{*}{${\bf DCD}$}
      & \multirow{5}{*}{$(\bf{b},\bf{a},\bf{b})$}
	  & $(\bf{1},-\bf{1},-\bf{1})$
	      & $(-\bf{1},-\bf{1},\bf{1})$
		  & \hfil $2$ & \hfil $2$ & \hfil $2$ & \hfil $2$ \\
  \cline{4-9}
  &&
	  & $(\bf{1},-\bf{1},-\bf{1})$
	      & $(-\bf{1},\bf{a},-\bf{b})$
		  & \hfil $2$ & \hfil $2$ & \hfil $2$ & \hfil $2$ \\
  \cline{4-9}
  &&
	  & $(-\bf{1},\bf{1},-\bf{1})$
	      & $(\bf{b},-\bf{1},-\bf{b})$
		  & \hfil $2$ & \hfil $2$ & \hfil $2$ & \hfil $8$ \\
  \cline{4-9}
  &&
	  & $(-\bf{1},\bf{a},-\bf{b})$
	      & $(\bf{b},-\bf{a},-\bf{1})$
		  & \hfil $2$ & \hfil $2$ & \hfil $2$ & \hfil $8$ \\
  \cline{4-9}
  &&
	  & $(-\bf{1},-\bf{a},\bf{b})$
	      & $(-\bf{b},\bf{a},-\bf{1})$
		  & \hfil $2$ & \hfil $2$ & \hfil $2$ & \hfil $2$ \\
  \cline{2-9}
  & \multirow{1}{*}{$\mathbf{(DDD)^{\prime}}$}
      & \multirow{1}{*}{$(\mathbf{b},\mathbf{b},\mathbf{b})$}
	  & $(\mathbf{1},-\mathbf{1},-\mathbf{1})$
	      & $(-\mathbf{1},-\mathbf{1},\mathbf{1})$
		  & \hfil $1$ & \hfil $1$ & \hfil $1$ & \hfil $1$ \\		  
  \hline
 \end{tabular}
\caption{\it $\Z_2 \times \Z_2$ orbifold models on the $A_3 \times A_3$ Lie
root lattice.}
\label{z2z2a3a3}
\end{center}
\end{table}

\section{Comments on lattices}
\label{App-lattice}

In this appendix we briefly summarize conventions of the (sub-)lattice and
its dual lattice space for a $\Z_2$ action $\cR$ in the following way:
\begin{eqnarray*}
\Lambda_{\cR,\perp}&:& \text{lattice projected out by the action $\cR$},~
\Lambda_{\cR,\perp} \equiv \frac{1+\cR}{2}\Lambda \\
\Lambda_{\cR,\inv}&:& \text{$\cR$ invariant sublattice} \\
\Lambda^*&:& \text{dual lattice of $\Lambda$, 
for its base
${\bf \alpha}_j \cdot {\bf \alpha}_i^* =\delta_{ji},~
{\bf \alpha}_j \in \Lambda,~{\bf \alpha}^*_i \in \Lambda^*$} 
\end{eqnarray*}
These three lattice spaces are closely related to one another.
Introducing a lattice $\Lambda_{-\cR, \perp}$ 
which is projected out by the $-\cR$ action on it, then 
we find the following non-trivial equations: 
\bsubeq
\bea
\Lambda^*_{\cR,\perp} &=& (\Lambda_{\cR,\inv})^*,\\
\vol(\Lambda) &=& \vol(\Lambda_{\cR,\inv}) \cdot \vol(\Lambda_{-\cR,\perp}),\\
\vol(\Lambda^*) &=& \vol(\Lambda)^{-1}.
\eea
\esubeq
Let us analyze in a more concrete way.
For example, 
we consider the four-dimensional 
$D_4$ Lie root lattice $\Lambda$ and its dual lattice $\Lambda^*$ 
based on
\begin{align}
\Lambda :~ \left\{
\begin{array}{@{(}r@{,}r@{,}r@{,}r@{)}}
\phantom{,}1 & -1 & 0 & 0 \\
0 & 1 & -1 & 0 \\
0 & 0 & 1 & -1 \\
0 & 0 & 1 & 1 
\end{array} \right.
\ls \ \ 
\Lambda^* :~ \left\{
\begin{array}{@{(}r@{,}r@{,}r@{,}r@{)}}
\phantom{,}1 & \phantom{,}0 & \phantom{,}0 & 0 \\
1 & 1 & 0 & 0 \\
\half & \half & \half & - \half \\
\half & \half & \half & \half 
\end{array} \right.
\end{align}
and we give a ${\mathbb Z}_2$ action $\cR$ on the $D_4$ lattice as 
\beq
\cR = \text{diag}(1,1,-1,-1) .
\eeq
Then, we can obtain the basis vectors in the lattices
$\Lambda_{\cR,\perp}$, $\Lambda_{\cR,\inv}$,
$\Lambda_{\cR,\perp}^*$ and $\Lambda_{\cR,\inv}^*$ in the following forms:
\bsubeq
\begin{align}
&&
\Lambda_{\cR,\perp} :& \ \left\{
\begin{array}{l}
(1,-1,0,0) \\
(0,1,0,0) 
\end{array} \right.
&
\Lambda_{\cR,\inv} :& \ \left\{
\begin{array}{l}
(1,-1,0,0) \\
(0,2,0,0) 
\end{array} \right.
&& \\
&&
\Lambda^*_{\cR,\perp} :& \ \left\{
\begin{array}{l}
(1,0,0,0) \\
(\frac12,\frac12,0,0) 
\end{array} \right.
&
\Lambda^*_{\cR,\inv} :& \ \left\{
\begin{array}{l}
(1,0,0,0) \\
(1,1,0,0) 
\end{array} \right.
\end{align}
\esubeq
Thus we easily see the relation among various lattice spaces:
\beq
\begin{array}{lllll}
\Lambda_{\cR,\perp} & \stackrel{\perp}{\longleftarrow} 
& \Lambda & \stackrel{\inv}{\longrightarrow} & \Lambda_{\cR,\inv} \\
 \updownarrow *  & & \updownarrow *  & & \updownarrow * \\
\Lambda^*_{\cR,\inv} & \stackrel{\inv}{\longleftarrow} 
& \Lambda^* & \stackrel{\perp}{\longrightarrow} & \Lambda^*_{\cR,\perp} 
\end{array}
\eeq

\section{String one-loop amplitudes}
\label{App-1loopPF}

In this appendix we summarize descriptions 
of the string one-loop amplitudes whose topologies
are given by  
the Klein bottle, the annulus and the M\"obius strip in the loop channel 
\cite{Forste:2000hx, Blumenhagen:1999ev}.
These are  applied to discuss the RR-tadpole amplitudes in the main
part of this paper. 
Here we start from the forms\footnote{In this appendix we
  borrow quite useful conventions and equations 
in appendix A of \cite{Forste:2000hx}.}
in which the zero mode and
the oscillator modes are factorized:
\bsubeq \label{app-KAM}
\begin{align}
{\cal K} &=  
4c (1_{\text{RR}} -1_{\text{NSNS}} ) 
\int_0^\infty \frac{dt}{t^3}\Bigg(
    \frac{1}{4NM} \sum_{n_1,k_1  =0}^N \sum_{n_2,k_2 =0}^M 
{\cal K}^{\left(n_1,k_1\right)\left(n_2,k_2\right)} 
{\cal L}_{\cal K}^{\left(n_1,k_1\right)\left(n_2,k_2\right)} \Bigg) , \\
{\cal A} &=  
c (1_{\text{RR}} -1_{\text{NSNS}} ) \int_0^{\infty}
    \frac{dt}{t^3} \Bigg(\frac{1}{4NM}\sum_{n_1,k_1 = 
    0}^{N}\sum_{n_2,k_2=0}^M
    \sum_{(i_1,i_2)=(0,0)}^{(N-1,M-1)}
\hspace{-3mm} \tr \big(\gamma_{k_1k_2}^{\left(i_1, i_2\right)} \big)
    \tr \Big( \big(\gamma_{k_1k_2}^{\left(i_1-n_1, 
    i_2-n_2\right)} \big)^{-1} \Big) \nonumber\\
&\hspace{5cm} \times 
{\cal A}^{\left(n_1,k_1\right)\left(n_2,k_2\right)}
{\cal L}_{\cal A}^{(n_1,k_1)(n_2,k_2)(i_1,i_2)} \Bigg),\\
{\cal M} & =  
-c (1_{\text{RR}} -1_{\text{NSNS}} ) \int_0^{\infty}
    \frac{dt}{t^3} \Bigg(\frac{1}{4NM}\sum_{n_1,k_1 =
    0}^{N}\sum_{n_2,k_2=0}^M
    \sum_{(i_1,i_2)=(0,0)}^{(N-1,M-1)}
\hspace{-3mm} \tr \Big(
    \big(\gamma_{\Omega{\cal R}k_1k_2}^{\left(i_1, i_2
    \right)} \big)^{-1} \big(\gamma_{\Omega{\cal R}k_1k_2}^{\left(i_1-n_1,
    i_2-n_2\right)} \big)^{T} \Big) \nonumber\\
&\hspace{5.5cm} \times {\cal
    M}^{\left(n_1,k_1\right)\left(n_2,k_2\right)}{\cal L}_{\cal
    M}^{(n_1,k_1)(n_2,k_2)(i_1,i_2)}\Bigg), 
\end{align}
\esubeq
where the values
${\cK}^{(n_1,k_1)(n_2,k_2)}$, ${\cA}^{(n_1,k_1)(n_2, k_2)}$ and
${\cM}^{(n_1,k_1) (n_2,k_2)}$ denote oscillator contributions, 
and ${\cal L}$ indicates the zero mode contributions in the amplitudes.
They belong to the $\theta^{n_1}\phi^{n_2}$-twisted sector 
with $\theta^{k_1}\phi^{k_2}$-insertion in the amplitudes. 
The $\gamma^{(i_1,i_2)}$'s are the matrix representations 
of the orientifold action on the Chan-Paton factors \cite{Gimon:1996rq},
whose superscript $(i_1,i_2)$ labels the different types of D6-branes
on which the open string attaches. 
The location of the brane $(i_1,i_2)$ is defined by 
rotating brane $(0,0)$ by the action $\theta^{-i_1/2}\phi^{-i_2/2}$. 

\subsection{Contributions from zero modes}

The above one-loop amplitudes (\ref{app-KAM}) contain  
the zero mode contributions $\cL_{\cK,\cA,\cM}$ 
from the sum of the Kaluza-Klein momentum modes and the 
winding modes, which are expressed in such a way as  
\bsubeq \label{app-zero-L}
\begin{eqnarray}
{\cal L}_{\cal K}^{\left(n_1,k_1\right)\left(n_2,k_2\right)} & = &
\chi_{\cal K} ^{\left(n_1, k_1\right)\left(n_2,k_2\right)}\mbox{Tr}^{\left(n_1
  ,n_2\right)}_{\mbox{\scriptsize KK}+\mbox{\scriptsize
  W}}\left(\Omega {\cal 
    R} \theta ^{k_1}\phi ^{k_2} e^{-2\pi t\left( L_0
      +\bar{L}_0\right)} \right), \\
{\cal L}_{\cal A}^{\left(n_1, k_1\right)\left( n_2,
    k_2\right)\left(i_1,i_2\right) }& = &\chi_{\cal A}^{\left(n_1,
    k_1\right)\left( n_2, 
    k_2\right)\left(i_1,i_2\right) 
    }\mbox{Tr}^{\left(i_1,i_2\right),\left(i_1 -n_1, i_2 -
    n_2\right)}_{\mbox{\scriptsize KK}+\mbox{\scriptsize W}} \left(
    \theta ^{k_1}\phi ^{k_2} e^{-2\pi t
    L_0}\right),\\
{\cal L}_{\cal M}^{\left(n_1, k_1\right)\left( n_2,
    k_2\right)\left(i_1,i_2\right) }& = &\chi_{\cal M}^{\left(n_1,
    k_1\right)\left( n_2, 
    k_2\right)\left(i_1,i_2\right) 
    }\mbox{Tr}^{\left(i_1,i_2\right),\left(i_1 - n_1, i_2 -
    n_2\right)}_{\mbox{\scriptsize KK}+\mbox{\scriptsize W}} \left(
    \Omega {\cal R} \theta ^{k_1}\phi ^{k_2} e^{-2\pi t
    L_0}\right). 
\end{eqnarray}
\esubeq
Note that 
in the Klein bottle amplitude 
$\chi_{\cK}$ denotes the number of the corresponding fixed
points which are left invariant
under orientifold group actions ${\cal R}\theta^{k_1}\phi^{k_2}$. 
In the open string amplitudes $\chi_{\cA}$ gives the 
intersection number of the D-branes involved. 
 
When we consider string propagating in the torus $T^6 =
\bR^6/\Lambda$, the zero modes contributions ${\cal L}$
from the momentum modes ${\bf p} = \sum_i n_i {\bf p}_i$ and 
the winding modes ${\bf w} = m_i {\bf w}_i$
are given by
\beq
\cL \equiv \sum_{n_i} \exp \Big( -\delta \pi t n_i M_{ij} n_j \Big)
\cdot \sum_{m_i} 
\exp \Big(-\delta \pi t m_i W_{ij} m_j \Big) ,
\eeq
where $t$ is the modulus in the loop channel and 
$n_i$, $m_i \in \bZ$ are the quanta in the momentum modes and the winding
modes \cite{Blumenhagen:2004di}. 
Note that the matrices $M_{ij}$ and $W_{ij}$ are
given by the products of ${\bf p}_i$ and of ${\bf w}_i$ in such a way as
$M_{ij} = {\bf p}_i \cdot {\bf p}_j$, 
$W_{ij} = {\bf w}_i \cdot {\bf w}_j$;
we set $\delta = 1$ (the Klein bottle), 
$\delta = 2$ (the annulus and the M\"obius strip). 
Due to this, in two-dimensional torus $T^2 \subset T^6$,
we can rewrite the above equations (\ref{app-zero-L})
in the following form:
\begin{equation}
{\cal L}_{ {\alpha} , \beta} \equiv
\sum_{m \in \Z}
  \exp \Big( - \frac{{\alpha} \pi t m^2}{\rho} \Big)
\cdot \sum_{n \in \Z} \exp \Big( -\beta \pi t n^2 \rho \Big) ,
\label{zero-loop}  
\end{equation}
where $\rho = r^2 / {\alpha}^\prime$.
It is worth rewriting this to the one in the tree channel.
According to the Poisson resummation formula
\begin{equation}
\sum_{n\in \Z} e^{-\pi n^2 /t} = \sqrt{t}\sum_{n\in \Z} e^{-\pi n^2 t} ,
\end{equation}
we find that the zero mode contribution
 in the tree channel is given as 
\begin{equation}
\tilde{{\cal L}}_{ {\alpha} , \beta} \equiv 
\sum_{m \in \Z}
  \exp \Big( -{\alpha} \pi l m^2  \rho \Big) 
\cdot \sum_{n \in \Z} \exp \Big( - \frac{\beta \pi l n^2}{\rho} \Big) .
\label{zero-tree} 
\end{equation}
This formulation is quite useful not only for factorizable torus $T^2
\times T^2 \times T^2$ but also for non-factorizable tori in the main
text via a suitable arrangement.

\subsection{Contributions from oscillator modes}

Here we move to the discussion on the oscillator modes.
These contributions into the one-loop amplitudes (\ref{app-KAM}) are given by 
\bsubeq \label{nknk}
\begin{align}
{\cal K}^{(n_1,k_1)(n_2,k_2)} 
&=  
\Tr^{(n_1 ,n_2)}_{\text{NSNS}} 
\Big(\Omega {\cal R} \theta^{k_1} \phi^{k_2} (-1)^F 
e^{-2\pi t ( L_0 +\bar{L}_0 )} \Big) 
, \label{K-nknk} \\ 
{\cal A}^{( n_1 , k_1 ) (n_2 ,k_2 )} 
&=
  \Tr^{( 0 , 0 ) (- n_1 , - n_2 )}_{\text{NS}}
\Big( \theta^{k_1} \phi^{k_2} (-1)^F 
e^{-2\pi t L_0} \Big)
, \label{A-nknk} \\ 
{\cal M}^{\left(n_1,k_1\right)\left(n_2, k_2\right)}
&= \mbox{Tr}_{\mbox{\scriptsize
    R}}^{\left(0,0\right) \left(-n_1,- 
  n_2\right)}\left(\Omega {\cal R}\theta ^{k_1}\phi ^{k_2}
e^{-2\pi tL_0}\right)
. \label{M-nknk}
\end{align}
\esubeq
The superscript $(0,0)(-n_1, -n_2)$
on the trace $\Tr_{\text{NS}}^{(0,0)(-n_1,-n_2)}$ in (\ref{A-nknk})
indicates open string states stretching between two distinct branes 
$(0, 0)$ and $(-n_1, -n_2)$, or equivalently, between
the brane $( i_1 , i_2 )$ and the brane 
$( i_1 -n_1 , i_2 - n_2 )$. 
The oscillator contributions (\ref{nknk}) 
can be expressed by the use of Jacobi theta
functions $\vartheta \left[ \alpha \atop \beta \right] (t)$ and the Dedekind eta
function $\eta (t)$:
\begin{align}
\vartheta\left[ {\alpha} \atop \beta \right] \left(
  t\right) 
  &=  \sum_{n\in \Z} q^{\frac{\left( n+{\alpha}\right)^2}{2}}e^{2\pi
  i\left( n+{\alpha}\right)\beta} , \ls
\eta\left( t\right) = q^{\frac{1}{24}}\prod_{n=1}^{\infty} \left(
  1 - q^n\right) ,
\label{eta-theta}
\end{align}
with $q = e^{-2\pi t}$. 
Then the amplitudes are expressed as  
\bsubeq \label{theta-eta}
\begin{align}
\cK^{( n_1 , n_2)} 
&= 
\frac{\vartheta\left[ 0 \atop 1/2 \right]}{\eta^3}
\prod_{n_1v_i + n_2w_i\not{\in} \Z}
\left( \frac{ \vartheta
\left[ n_1v_i + n_2w_i \atop 1/2 \right]}{\vartheta
\left[ 1/2+n_1v_i +n_2w_i \atop 1/2 \right]} 
e^{\pi i\langle n_1v_i + n_2w_i\rangle} \right) 
\nn \\
&\LS \times 
\prod_{n_1v_i +n_2w_i\in \Z}
\left(\frac{\vartheta \left[ 0 \atop 1/2 \right]}{\eta^3}\right) 
, \label{K-theta-eta} \\
\cA^{( n_1 ,k_1 ) ( n_2 ,k_2 )} 
&= 
\frac{\vartheta
\left[ 0 \atop 1/2 \right]}{\eta^3}
\prod_{(n_1v_i + n_2w_i,k_1v_i +k_2w_i)\not{\in} \Z^2}
\left( \frac{( -2i )^\delta  
\vartheta
\left[ n_1v_i + n_2w_i \atop 1/2 + k_1v_i+k_2w_i \right]}{\vartheta
\left[ 1/2+n_1v_i +n_2w_i \atop 1/2+k_1v_i +k_2w_i \right]}
        e^{\pi i\langle n_1v_i + n_2w_i\rangle}\right)
\nn \\
&\LS \times
\prod_{(n_1v_i +n_2w_i,k_1v_i +k_2w_i )\in \Z^2}
\left(\frac{\vartheta \left[ 0 \atop 1/2 \right]}{\eta^3}\right)
, \label{A-theta-eta} \\ 
{\cal M}^{(n_1 ,k_1)(n_2 ,k_2)} 
&=
\frac{\vartheta \left[ 1/2 \atop 0 \right]}{\eta^3}
\prod_{(n_1v_i + n_2w_i,k_1v_i +k_2w_i) \not{\in} \Z^2}
\left( \frac{(-2i)^\delta \vartheta
\left[ 1/2+n_1v_i + n_2w_i \atop k_1v_i+k_2w_i \right]}{\vartheta
\left[ 1/2+n_1v_i +n_2w_i \atop 1/2+k_1v_i +k_2w_i \right]}
        e^{\pi i\langle 
        n_1v_i + n_2w_i\rangle}\right)
\nn \\
&\LS \times
\prod_{(n_1v_i +n_2w_i,k_1v_i +k_2w_i)\in \Z^2}
\left(\frac{\vartheta
\left[ 1/2 \atop 0 \right]}{\eta^3}\right)
. \label{M-theta-eta}
\end{align}
\esubeq
Notice that except for the $\Z'_6$ orbifold the values 
$\cK^{(n_1,k_1)(n_2,k_2)}$ are equal for any insertion of
$\theta^{k_1} \phi^{k_2}$, even though the lattice contributions
differ \cite{Blumenhagen:1999ev}.
Then we omit the label $k_i$ in (\ref{K-theta-eta}).
The arguments in the theta and eta functions are $2t$ in the Klein
bottle, $t +\frac{i}{2}$ in the M\"obius strip, and $t$ in the annulus.
Further, we used the notation\cite{Blumenhagen:1999ev},
$\langle x\rangle \equiv x - \left[ x\right] - \frac{1}{2}$, 
where the brackets on the rhs denote the integer part and
\begin{equation}
\delta =\left\{ \begin{array}{c@{\ls}l}
1 & \text{if \ \ $\left(n_1v_i + n_2w_i, k_1 v_i + k_2 w_i\right) 
\in \Z \times \Z+\frac{1}{2}$} \\
0 & \text{otherwise} \end{array}\right.
\end{equation}
The tree channel expressions $\ctK$,
$\ctA$ and $\ctM$ can be evaluated with
the help of the modular transformation of (\ref{eta-theta}).


}
\end{document}